\documentclass[reprint,aps,prd,notitlepage,nofootinbib,floatfix,superscriptaddress, preprintnumbers]{revtex4-2}

\usepackage[utf8]{inputenc} 
\usepackage[T1]{fontenc}    
\usepackage{lmodern}
\usepackage[colorlinks=true,linkcolor=blue,citecolor=teal,urlcolor=teal,plainpages=false,pdfpagelabels]{hyperref}       
\usepackage{url}            
\usepackage{amsfonts}       
\usepackage{amsmath,amsthm,amssymb} 
\usepackage{bm}
\usepackage{bbm}
\usepackage{dcolumn}
\usepackage{nicefrac}       
\usepackage{graphicx}
\usepackage{xcolor, etoolbox}
\usepackage{mathtools}
\usepackage{mathrsfs}
\usepackage{soul}
\usepackage{ragged2e}
\usepackage{microtype}      
\usepackage{braket}      
\usepackage{enumitem}
\usepackage{xspace}
\usepackage{dsfont}
\usepackage{float}
\usepackage[export]{adjustbox}
\usepackage{lipsum} 
\usepackage{cleveref} 
\usepackage{overpic}
\usepackage{comment}

\newcommand{\bfk}{\bm{k}}
\newcommand{\bfx}{\bm{x}}

\newcommand{\dd}{\text{d}}
\newcommand{\vk}{\bm{k}}
\newcommand{\tk}{k}

\begin{document}
	
\title{Relevance of `on' and `off' transitions in quantum pair production experiments}

\author{\'{A}lvaro \'{A}lvarez-Dom\'{\i}nguez}
\email{alvalv04@ucm.es}
\affiliation{Departamento de Física Te\'{o}rica and IPARCOS, \\ Universidad Complutense de Madrid, 
Plaza de las Ciencias 1, 
28040 Madrid, Spain}

\author{\'{A}lvaro Parra-L\'{o}pez}
\email{alvaro.parra@ift.csic.es}
\affiliation{Departamento de Física Te\'{o}rica and IPARCOS, \\ Universidad Complutense de Madrid, 
Plaza de las Ciencias 1, 
28040 Madrid, Spain}
\affiliation{Instituto de F\'isica Te\'orica UAM/CSIC, calle Nicol\'as Cabrera 13-15, \\ Universidad Autónoma de Madrid, Cantoblanco, 28049 Madrid, Spain}

\preprint{IPARCOS-UCM-25-025}
\preprint{IFT-UAM/CSIC-25-44}

\flushbottom

\begin{abstract}
Analog gravity experiments, such as those realized in Bose-Einstein condensates,  often aim at simulating cosmological pair production due to the dynamical expansion of the Universe. However, these experiments have a start and an end, which introduces unavoidable transitions out of and into static regimes that alter the intended expansion profile. We show that the resulting particle spectra can be overwhelmingly dominated by these transition periods, which calls for a careful interpretation of experimental outcomes. In prospective Schwinger effect experiments, by contrast, transition effects do not dominate particle production, and such a reinterpretation may not be necessary.
\end{abstract}

\maketitle

\section{Introduction}
\label{sec:intro}

Particle creation phenomena are naturally described by quantum field theories in nontrivial backgrounds. Prominent examples include cosmological particle production resulting from the expansion of the Universe~\cite{Parker1968,Parker1969,Ford1987}, Hawking radiation emitted by black holes~\cite{Hawking1974,Hawking1975}, and the Schwinger effect, which occurs in the presence of strong electromagnetic fields~\cite{Sauter1931,Heisenberg1936,Schwinger1951}.

These scenarios are typically difficult to access experimentally. Instead, motivated by the original idea by Unruh~\cite{Unruh1981}, analog gravity experiments \cite{Barcelo2011,Jacquet2020,Almeida2023} have been used as a tool to explore the dynamics of quantum fields in nontrivial backgrounds or effective curved spacetimes. In recent years, numerous experiments have been carried out in hydrodynamical, condensed matter, optical systems, and others \cite{Philbin2008,Weinfurtner2011,Euve2016,MunozDeNova2019,Drori2019,Shi2023,Torres2017,Eckel2018,Wittemer2019,Banik2021,Giacomelli2021,Braidotti2022,Steinhauer2022,Jacquet2022,Viermann2022}, demonstrating the potential of such platforms for the study of quantum fields.

Cosmological analog experiments often aim to measure the analog of particle production caused by the expansion of the Universe over a specific cosmological time interval~\cite{Wittemer2019,Banik2021,Steinhauer2022,Viermann2022,Sparn2024}. Similarly, in Schwinger effect experiments, one might wonder what is the production of particles due to an electric field that is switched on during a certain period of time \cite{Schuetzhold2009,Bulanov2010,Aleksandrov2022,Aleksandrov2025,Ilderton2022}. However, in all these situations, one cannot avoid the existence of transitions from and to static regimes in which the cosmological expansion ceases or the electric field vanishes. Analog experiments have a beginning and an end, and electric fields must be switched on and off to implement specific pair production processes in the laboratory. Therefore, a very natural question arises: How do these `on' and `off' transitions impact the results of the experiments? Are they negligible, or do they affect the particle production process? If the latter is true, one has to be careful when interpreting the results of such experiments, as the particle production occurring during the particular time window that one is trying to simulate could be overshadowed by the production taking place during the transitions.

To address these questions, we first derive fundamental insights from the more general case of cosmological pair production in homogeneous and isotropic cosmologies. Indeed, these transition regimes are also present in early universe scenarios, where the computation of cosmologically produced particles typically relies on the fact that the Universe's expansion becomes sufficiently slow at very early and late times---such as at the onset of inflation and well into the reheating epoch, respectively. Our analysis is therefore also interesting in actual cosmological scenarios, and characterizes which regions of spacetime are more relevant regarding cosmological pair production.

Specifically, we study the impact of `on' and `off' transitions on particle production in $D$-dimensional Friedmann-Lemaître-Robertson-Walker (FLRW) expanding universes. We will consider a massive spectator scalar field, allowing for a nonminimal coupling to the geometry. Our results demonstrate that the coupling parameter between the field and the geometry has a strong influence on the resulting pair production spectrum. In particular, we find that particle production during the targeted time window is inevitably affected by `on' and `off' transition periods. Furthermore, in the case of a nonconformal coupling, production during abrupt transitions dominates pair creation, significantly overshadowing the contributions from the intermediate region. However, when the coupling is conformal, the `on' and `off' transitions do not substantially enhance pair creation, resulting in a much lower overall density of produced particles.

We will then apply these fundamental results to two experimental setups, highlighting the need for extreme caution when interpreting the physical results of these experiments. On the one hand, we will discuss analog pair production in (1+2)-dimensional Bose-Einstein condensates (BECs), which simulates the problem of a nonconformally coupled field in an FLRW universe. On the other hand, we will discuss the Schwinger effect due to a switchable electric field in $(1+3)$ dimensions. Its behavior regarding `on' and `off' transitions is for the most part equivalent to that of a conformally coupled field in an FLRW universe. However, unlike the cosmological case, the contribution from the intermediate regime increasingly dominates over that of the abrupt transitions when the field remains switched on for a sufficiently long time, leading to enhanced particle production as the field duration grows.

This paper is as follows. In Sec.~\ref{sec:cosmo}, we review cosmological pair production for a scalar field in FLRW in $D$ spatial dimensions. In Sec.~\ref{sec:switchonoff}, we analyze the impact of the transitions between static and dynamic regimes of the scale factor on particle production. Then, in Sec.~\ref{sec:BEC}, we apply these ideas to the case of analog pair production in BECs. We discuss these matters in the context of the Schwinger effect in Sec.~\ref{sec:Schwinger}. Finally, we elaborate our conclusions in Sec.~\ref{sec:conclusions}. 

We work in natural units, setting $c = \hbar = \varepsilon_0 = 1$.

\section{Cosmological pair production}
\label{sec:cosmo}

Let us consider a $(1+D)$-dimensional FLRW spacetime with vanishing spatial curvature \cite{Friedman1922, Friedman1924, Lemaitre1931, Robertson1935, Robertson1936a, Robertson1936b, Walker1937}, 
\begin{equation}
\dd s^2 = a^2(\eta)\left(-\text{d} \eta^2+\text{d} \bfx^2 \right),
 \label{eq:GeneralLineElement}
\end{equation}
where~$\eta$ is the conformal time and $a(\eta)$ the scale factor. The dynamics of a real, nonminimally coupled to gravity scalar field~$\varphi(\eta,\bfx)$ with mass~$m$ is described by the equation
\begin{equation}
    \varphi^{\prime\prime} + (D-1)\mathcal{H}\varphi^\prime - a^2(\Delta + m^2 + \xi R) \varphi = 0,
\end{equation}
where ${}^\prime = \partial/\partial\eta$, $\Delta$ is the Laplace operator and $\mathcal{H}=a^\prime / a$ the conformal Hubble parameter. The field is coupled via the parameter~$\xi$ to the Ricci curvature scalar~$R$.

It is convenient to work with the rescaled field \mbox{$\chi =  a^{(D-1)/2} \varphi$}, whose Fourier modes $\chi_{\vk}$ satisfy a decoupled system of ordinary differential equations of the form 
\begin{equation}
    \chi^{\prime\prime}_{\vk}(\eta) + \omega_{\tk}^2(\eta) \chi_{\vk}(\eta) = 0.
    \label{eq:ModeEquation}
\end{equation}
These are harmonic oscillator equations with time-dependent frequencies, given by
\begin{align}
    \omega_{\tk}^2 &= \tk^2 + m^2a^2 \nonumber \\ &+\frac{1+(4\xi-1)D}{4} \left[2 \frac{a''}{a}+(D-3) \left( \frac{a'}{a} \right)^2 \right],
    \label{eq:MasterFrequency}
\end{align}
where $\tk=|\vk|$. These frequencies capture all the information about the gravitational background, which we consider external, classical, and not affected by the dynamics of the matter field~$\varphi$.

To canonically quantize the field~$\varphi$, we perform an expansion in terms of a basis of solutions $\{v_k,v^*_k\}$ to the mode equations~\eqref{eq:ModeEquation}, and promote the linear coefficients to annihilation and creation operators, $\hat{b}_{\vk}$ and~$\hat{b}_{\vk}^{\dagger}$:
\begin{align}
\hat{\varphi}(\eta,\bfx) &= a(\eta)^{\frac{1-D}{2}} \nonumber \\
\times &\int \frac{\text{d}^D\bfk}{\left(2\pi\right)^{\frac{D}{2}}} \ \left[\hat{b}_{\vk} v_k(\eta)e^{i\bfk\cdot\bfx}
+ \hat{b}_{\vk}^\dagger v_k^*(\eta)e^{-i\bfk\cdot\bfx}\right].
\label{eq:FieldExpansion}
\end{align}
The quantum vacuum is the state annihilated by all the annihilation operators, i.e.,~$\hat{b}_{\vk} |0\rangle=0$ for all~$\vk$. The operators satisfy the standard commutation relations \mbox{$[\hat{b}_{\vk},\hat{b}_{\vk'}^{\dagger}]=\delta(\vk-\vk')$}, whereas all the other commutators vanish. 

In the following, we consider an initially static universe that begins expanding at some finite time~$\eta_{\text{on}}$ until it halts expansion and returns to a static state from a later time~$\eta_{\text{off}}$. This scenario is sensible in the context of the inflationary universe, at the beginning of which the geometry expands slowly. After inflation, the universe thermalizes, and the expansion becomes again very adiabatic. This defines `in' and `out' regions in which the universe is static, and the frequency \eqref{eq:MasterFrequency} becomes constant. Since we have an initial period of staticity, a natural choice of basis is the set of solutions to Eq.~\eqref{eq:ModeEquation} that behave as positive-frequency plane waves before $\eta_{\text{on}}$, i.e.,
\begin{equation}
    v^{\text{in}}_k(\eta)=(2\omega_k^{\text{in}})^{-\frac{1}{2}}e^{-i\omega_k^{\text{in}}\eta}, \quad \eta\leq \eta_{\text{on}},
\label{eq:vin}
\end{equation}
with $\omega_k^{\text{in}}=\sqrt{k^2+ m^2a^2(\eta_{\text{on}})}$. This choice leads to the so-called `in' vacuum,~$|0^{\text{in}}\rangle$. Analogously, the `out' vacuum is given by the solutions that behave after the expansion as plane waves with frequencies \mbox{$\omega_k^{\text{out}}=\sqrt{k^2+m^2a^2(\eta_{\text{off}})}$}:
\begin{equation}
    v^{\text{out}}_k(\eta)=(2\omega_k^{\text{out}})^{-\frac{1}{2}}e^{-i\omega_k^{\text{out}}\eta}, \quad \eta\geq \eta_{\text{off}}.
\label{eq:vout}
\end{equation}

The `in' and `out' bases are associated with the natural quantizations of observers living before and after the expansion of the universe. Due to the time-dependence of the geometry, even if the initial state of the system is the vacuum state as understood by the observer at $\eta\leq\eta_{\text{on}}$, it will be in general an excited state for another observer at $\eta\geq\eta_{\text{off}}$, where the particle content is nonvanishing. To measure this particle creation, we write the `out' modes in terms of the `in' modes,
\begin{equation}
v^{\text{out}}_k=\alpha_kv^{\text{in}}_k+\beta_kv_k^{\text{in}*},
\label{eq:Bogoliubov}
\end{equation}
where $\alpha_k$ and $\beta_k$ are Bogoliubov coefficients \cite{Birrell1982,Mukhanov2007}, which satisfy \mbox{$|\alpha_k|^2 - |\beta_k|^2 = 1$}. The expectation number density of excited quanta per mode~$k$ is precisely given by the square modulus of the $\beta$ coefficient,
\begin{equation}
    n_k = \langle 0^{\text{in}}| \hat{b}_k^{\text{out}\dagger} \hat{b}_k^{\text{out}} |0^{\text{in}}\rangle = |\beta_k|^2,
    \label{eq:nk}
\end{equation}
where this coefficient is time independent and can be computed in terms of the `in' and `out' solutions,
\begin{equation}
    \beta_k = i\left\{ v^{\text{in}}_k(\eta) \left[v^{\text{out}}_k(\eta)\right]^\prime - v^{\text{out}}_k(\eta) \left[v^{\text{in}}_k(\eta)\right]^\prime \right\}.
    \label{eq:betakvinvout}
\end{equation}
The total number density of produced pairs is given by the sum to all modes:
\begin{equation}
    n = \int \text{d}^D\bfk \ n_k.
    \label{eq:n}
\end{equation}

\section{`on' and `off' transitions}
\label{sec:switchonoff}

We aim to model the expansion of the Universe occurring between two times, $\eta_{\text{on}}$ and~$\eta_{\text{off}}$. However, the specific way in which we model the transition between the `in' regime ($\eta \leq \eta_{\text{on}}$) and the intermediate region ($\eta_{\text{on}} \leq \eta \leq \eta_{\text{off}}$), as well as the transition between this intermediate region and the `out' regime ($\eta \geq \eta_{\text{off}}$), inevitably influences particle production. Whether these transitions are abrupt or adiabatic can greatly affect production. This raises a critical question: To what extent can the impact of these transitions be considered negligible? Are there cases where particle production during these phases becomes so pronounced that it masks the effects of the expansion we intend to simulate? Our analysis shows that transition effects are inevitable in all cases and have a significant impact on pair creation.

\begin{figure}
    \centering
    \includegraphics[width=0.375\textwidth]{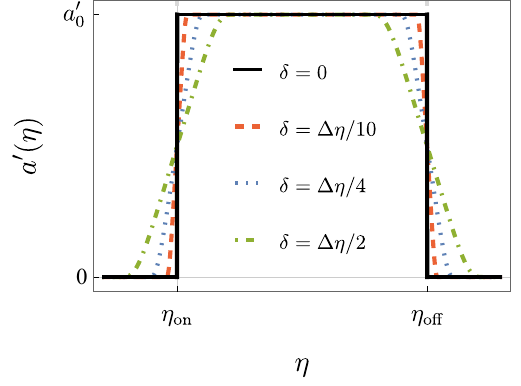}
    \caption{Expansion rate in Eq.~\eqref{eq:DerivativeScale} as function of conformal time, for different transition durations $\delta$, with $\Delta\eta = \eta_{\text{off}} - \eta_{\text{on}}$.}
    \label{fig:SwitchOnAndOff}
\end{figure}

Our goal is to identify whether particle production during the transitions is more or less significant than the production during the intermediate expansion---the regime in which we are actually interested in the case of analog experiments, for example. To illustrate these ideas, we consider a universe undergoing a constant expansion rate in terms of conformal time; i.e., \mbox{$a^\prime(\eta)=a_0^\prime$}, for~\mbox{$\eta_{\text{on}} \leq \eta \leq \eta_{\text{off}}$}. We model the transitions between this intermediate region and the static regimes by the regularized interpolation function 
\begin{equation}
    \Theta_{\delta}(\eta)= \left(1+\tanh \left\{\cot\left[\pi\left(1/2-\eta/\delta \right)\right]\right\}\right)/2,
    \label{eq:Interpolation}
\end{equation}
for $-\delta/2\leq \eta \leq \delta/2$, remaining constant outside this interval. $\Theta_{\delta}(\eta)$ is centered at the origin and its width is parametrized by~$\delta$, such that for~$\delta=0$ we recover the discontinuous Heaviside step function. We can then write the expansion rate during the entire expansion, including the transitions, as
\begin{equation} 
\label{eq:DerivativeScale}
    a^{\prime}(\eta)=a_0^\prime\left[\Theta_{\delta}(\eta-\eta_{\text{on}}-\delta/2)-\Theta_{\delta}(\eta-\eta_{\text{off}}+\delta/2)\right].
\end{equation}
In Fig.~\ref{fig:SwitchOnAndOff}, we represent this expansion rate for various transition durations~$\delta$. We restrict our analysis to transitions that are fast compared to the characteristic expansion rate in the intermediate region---for the specific scale factor in \eqref{eq:DerivativeScale}, this corresponds to $\delta a_0’ \lesssim 1$. In the following, we will say that the scale factor undergoes an \textit{abrupt transition} when it is continuous but not differentiable. During these abrupt transitions, the expansion rate~$a^\prime(\eta)$ involves discontinuous (but finite) step functions, as depicted in Fig. \ref{fig:SwitchOnAndOff} for~$\delta=0$. In the subsequent figures, we fix the intermediate expansion rate to $a_0^\prime = m$.

\begin{figure*}
    \centering
    \includegraphics[width=\textwidth]{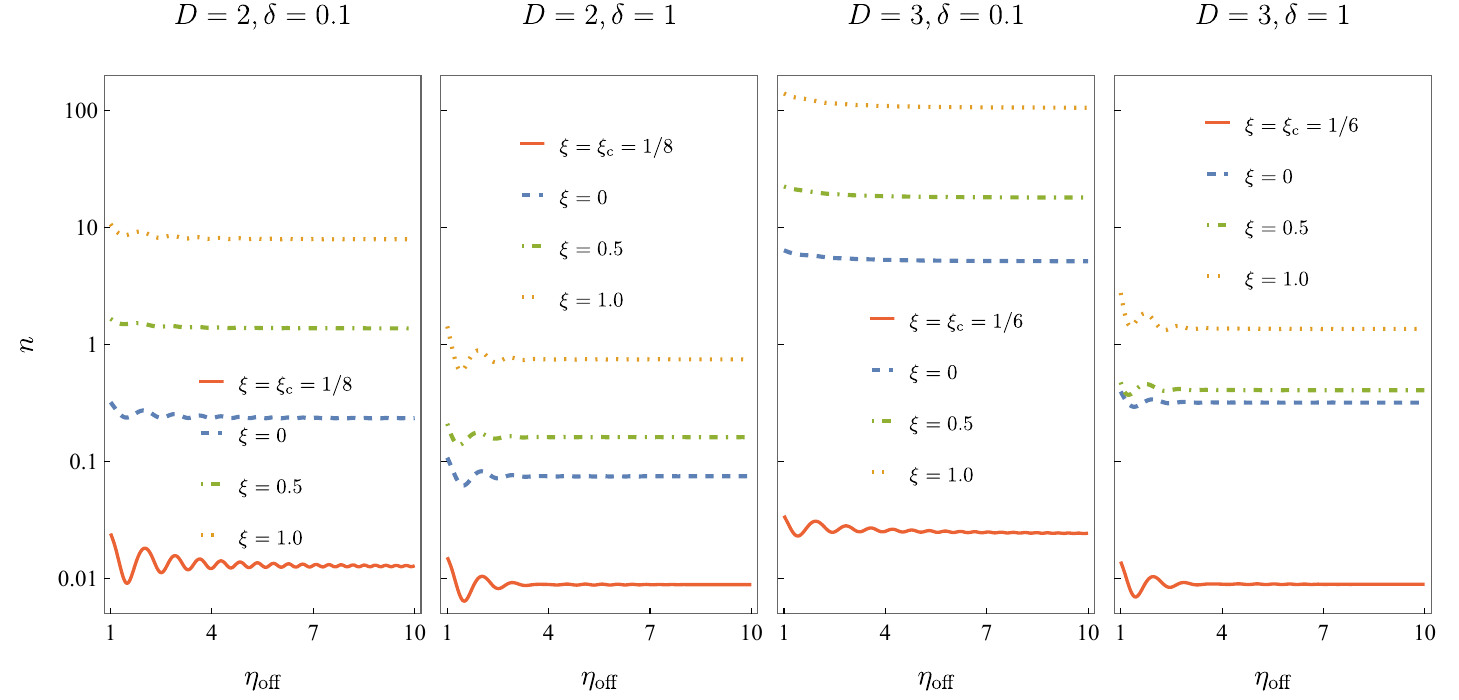}
    \caption{Total number density of produced particles $n$ as a function of $\eta_{\text{off}}$ ($\eta_{\text{on}}=0$), in the case of two and three spatial dimensions. Results are shown for different values of the coupling~$\xi$, where the red continuous lines correspond to the conformal coupling case. `On' and `off' transitions are chosen sufficiently fast ($\delta=0.1$) to guarantee convergence to the limiting behavior as the product $\delta a_0^\prime$ becomes small. We also consider slower transition rates, comparable to the expansion rate in the intermediate region ($\delta=1$). Here, $\delta$ and $\eta_{\text{off}}$ are expressed in units of $m^{-1}$, while the number density $n$ is given in units of $m^3$.}
    \label{fig:TotalDensity}
\end{figure*}

In Fig.~\ref{fig:TotalDensity}, we calculate the total number density of produced pairs, as defined in Eq.~\eqref{eq:n}, for various durations of the intermediate expansion, $\eta_{\text{on}} \leq \eta \leq \eta_{\text{off}}$, and for different transition durations~$\delta$. Specifically, we fix $\eta_{\text{on}} = 0$ and numerically compute the $\beta$-Bogoliubov coefficient using Eq.~\eqref{eq:betakvinvout} for each process, with its duration parametrized by a particular value of~$\eta_{\text{off}}$. We then integrate over all modes~$k$ to obtain the total density. This procedure is repeated for each value of~$\eta_{\text{off}}$. We examine how these particle densities depend on the coupling parameter $\xi$ and present results for $D=2$ and $D=3$ spatial dimensions. Our analysis reveals that 1) the total number density of produced pairs in the conformal coupling case, where \mbox{$\xi=(D-1)/(4D)$}, is significantly lower---by several orders of magnitude---than in the nonconformal case; \mbox{2) particle} production stabilizes for sufficiently prolonged expansions; and 3) for nonconformal couplings, fast transitions yield a much larger asymptotic number density than slower transitions.

\begin{figure*}
    \centering
    \includegraphics[width=0.7\textwidth]{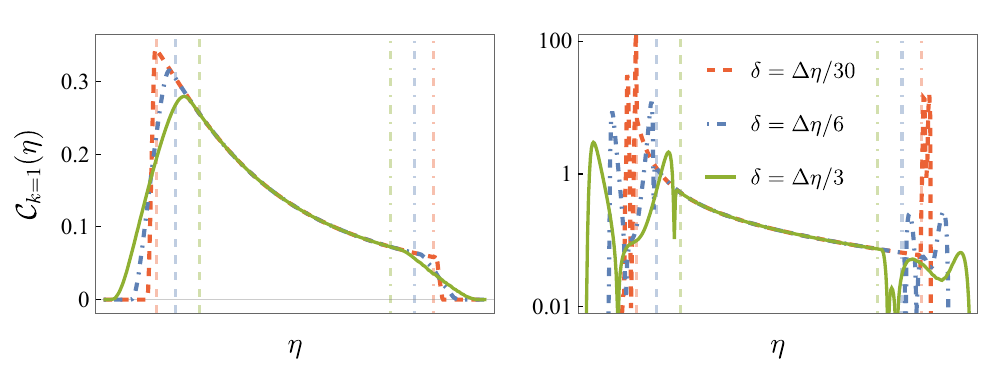}
    \caption{Function $\mathcal{C}_k$ for $D=3$ dimensional universe expansion with \mbox{$\Delta \eta=3$}, for different transition durations~$\delta$, fixed \mbox{$k=1$}. The left panel illustrates the conformal coupling case ($\xi=1/6$) while the right panel corresponds to a nonconformal coupling case ($\xi= 1$). Here, $\eta$ and $k$ are expressed in units of $m^{-1}$ and $m$, respectively.}
    \label{fig:Ck}
\end{figure*}

\textit{1) Production in the conformal case is suppressed with respect to the nonconformal case.}
The difference between these two cases can be understood by examining the behavior of the frequency \eqref{eq:MasterFrequency}, since pair production is dictated by the dynamics of the field modes. It is when the frequency rapidly varies over time, that particle production is enhanced. During the intermediate expansion phase ($\eta_{\text{on}} \leq \eta \leq \eta_{\text{off}}$), the frequency remains bounded in both cases, provided the expansion is sufficiently smooth. However, a crucial distinction arises during the abrupt transitions: while~$a^\prime$ remains bounded, $a^{\prime\prime}$ approaches a Dirac delta. In the conformal coupling case, the mode frequency simplifies to~\mbox{$\omega_k^2 = k^2 + m^2a^2$}, meaning that both $\omega_k$ and its time derivative remain bounded even during abrupt transitions. Conversely, for nonconformal coupling, the frequency explicitly depends on the first and second derivatives of the scale factor. As a result, the frequency and its time derivatives sharply diverge at the transition points, leading to significant enhancement of particle production. The crucial point is that this enhancement arises not from the intermediate expansion, but rather from the `on' and `off' transitions. This is particularly evident in our example, as in our computations we employ a constant expansion rate during the intermediate regime, ensuring that the derivatives of the scale factor entering the frequency~\eqref{eq:MasterFrequency} vanish except at the transitions\footnote{Although the quantitative differences between conformal and nonconformal couplings may vary depending on the specific form of the scale factor, the qualitative behavior highlighted here is robust.}.

 \textit{2) Production stabilizes for sufficiently prolonged intermediate expansions.} Even when the intermediate expansion lasts significantly longer, the particle number density remains effectively constant, indicating that, from a certain point on, the intermediate phase contributes minimally to the overall production. We will see that the Schwinger effect exhibits a drastically different asymptotic behavior.

\textit{3) For nonconformal couplings, fast transitions yield a much larger asymptotic number density than slower transitions.} This behavior reinforces the conclusion that the dominant contribution to particle creation arises from the abrupt `on' and `off' transitions. When the coupling is nonconformal, considering slower transitions in comparison to the expansion rate in the intermediate region significantly reduces production. The resulting horizontal asymptote in these cases can thus be interpreted as capturing mainly the average effect of the abrupt transitions for fast transitions. This effect is comparatively suppressed in the conformal coupling scenario, which tells us that transitions do not have a significant impact in this case, consistent with point 1).

In order to characterize the rate of change of the mode frequency (for a generic expansion rate), we define the dimensionless function
\begin{equation} 
    \mathcal{C}_k(\eta)=\left|\frac{\omega_k^{\prime}(\eta)}{\omega_k^2(\eta)}\right|.
    \label{eq:adCk}
\end{equation}
This provides a straightforward-to-compute quantity, as numerical methods are not required---unlike in the evaluation of the number density of produced particles. From the expression of $\omega_k$ in Eq.~\eqref{eq:MasterFrequency}, it follows that $\mathcal{C}_k$ is a strictly decreasing function of $k$, reflecting that the time variation of the frequency always decreases as one considers larger wave numbers. This behavior results in a suppression of particle creation in the ultraviolet. Regarding its dependence on the conformal time~$\eta$, the function~$\mathcal{C}_k(\eta)$ is directly linked to the fluctuations of the particle density per mode $k$, as captured by the Quantum Vlasov Equation~\cite{Kluger1998,Schmidt1998,Alvarez2022}. When $\mathcal{C}_k(\eta)$ becomes large during the transition phases, the particle number density experiences rapid oscillations, leading to enhanced particle production compared to the intermediate expansion period. On the other hand, if $\mathcal{C}_k(\eta)$ remains small during the transitions, the time variation of particle production is comparatively less oscillating, and the overall production rate is significantly lower. 

The frequency~\eqref{eq:MasterFrequency} depends in general on the derivatives of the scale factor, and 
\begin{align}
\mathcal{C}_k &=\omega_k^{-3} \left|m^2a a^{\prime} + \frac{1+(4\xi-1)D}{4}\right. \nonumber \\ &\left.\times \left[ \frac{a^{\prime\prime\prime}}{a} + (D-4)\frac{a^{\prime}a^{\prime\prime}}{a^2} - (D-3)\left( \frac{a^{\prime}}{a} \right)^3\right]\right|.
\label{eq:Cknonconformal}
\end{align}
For abrupt transitions, $a^{\prime\prime\prime}$ approaches the derivative of a Dirac delta, which strongly dominates over the terms proportional to $a'$ and $a^{\prime\prime}$. In the frequency~\eqref{eq:MasterFrequency}, the term with~$a^{\prime\prime}$ dominates. From Eq.~\eqref{eq:Cknonconformal}, this leads to $\mathcal{C}_k$ exploding during abrupt transitions, allowing production for a broad range of modes. Therefore, abrupt transitions in an expanding universe dramatically impact the spectra of produced particles, masking the contributions from the actual expansion process itself without such transitions.

In the particular case of conformal coupling, Eq. \eqref{eq:adCk} simplifies to
\begin{equation}
    \mathcal{C}_k=\left|\frac{m^2a a^{\prime}}{\left( k^2+m^2a^2 \right)^{3/2}}\right|,
    \label{eq:adCk16}
\end{equation}
which is bounded from above by $\mathcal{C}_{k=0}=\mathcal{H}/m$. Even in cases where the scale factor undergoes abrupt transitions, the function $\mathcal{C}_k$ remains finite, as it depends only on the first derivative of the scale factor and not on higher orders. Nevertheless, higher-order time derivatives of the frequency involve higher derivatives of the scale factor, which, in the limit of abrupt transitions, tend to Dirac delta distributions and their derivatives. Consequently, although still relevant, the impact of transitions in pair production is smaller in this case than in the nonconformal scenario.

For the particular shape given in Eq.~\eqref{eq:DerivativeScale}, Fig.~\ref{fig:Ck} shows the function~$\mathcal{C}_k$ for different values of~$\delta$ in $D=3$ dimensions. In the conformal coupling case $\xi = 1/6$, $\mathcal{C}_k$ remains bounded throughout the entire expansion, even during abrupt transitions. However, in the nonconformal coupling case (\mbox{$\xi = 1$}), $\mathcal{C}_k$ exhibits sharp oscillations during the `on' and `off' transitions. The amplitude of these oscillations increases by several orders of magnitude as the transitions become shorter, highlighting the sensitivity of the system to more rapid transitions. It is clear in this case that the primary contribution to the particle excitation number arises predominantly from the transitions, overshadowing the effects of the linear expansion in the intermediate region.

Note that in scenarios where the scale factor varies rapidly---particularly involving abrupt decelerations---such variations act as effective `on' and `off' transitions. Oscillatory or cyclic cosmologies, with alternating phases of expansion and contraction as discussed in Refs.~\cite{Schmidt2024,Sparn2024,Agullo:2024lry}, exemplify this behavior. In these cases, particle production is significant throughout the entire evolution.

In Sec.~\ref{sec:BEC} and Sec.~\ref{sec:Schwinger} we present two illustrative examples from laboratory settings where we apply the results just developed. The first involves gravitational analog experiments with BECs that mimic the dynamics of a nonconformally coupled scalar field in an FLRW expanding universe. The second focuses on the Schwinger effect, whose anisotropic nature introduces important nuances to our analysis that we discuss below.

\section{BEC analog experiment}
\label{sec:BEC}

We focus on the problem of analog particle production in a quasi-two-dimensional, spin-0 BEC~\cite{Tolosa2022,Viermann2022,Sparn2024,Schmidt2024}. Low-energy excitations on top of the condensate's ground state behave as a massless scalar field propagating in a curved spacetime defined by the so-called acoustic metric. This acoustic metric, determined by the properties of the condensate, can be experimentally controlled to emulate a two-dimensional FLRW metric. As a result, the system provides an analog for cosmological particle production in a $(1+2)$-dimensional spacetime.

The role of the scale factor in this analog setup is played by the scattering length, whose time dependence can be precisely controlled using Feshbach resonances~\cite{Stwalley1976,Cornish2000,Chin2010}. The expansion process is implemented across various stages: an initial `in' region where the scattering length remains constant, followed by an `on' transition into an intermediate region designed to mimic the desired cosmological scenario, and finally an `off' transition leading to a final `out' region where the scattering length returns to a constant value.

In analog experiments, the `on' and `off' transitions are unavoidable, and typically modeled as instantaneous. This approach was adopted in, e.g., Refs.~\cite{Viermann2022,Sparn2024}, where the abrupt transition model was shown to align well with experimental data. However, understanding the impact of these transitions on particle production is crucial. While the primary focus of such experiments lies in the intermediate region, where the desired scale factor behavior is replicated, it is essential to analyze how these transitions influence the dynamics to properly isolate and interpret the physical effects of interest.

In this analog platform, we do not have the freedom to select the value of $\xi$, which is zero in this case. For flat spatial sections, the mode equation corresponds to taking $D=2$ and \mbox{$m=0$} in the time-dependent frequency~\eqref{eq:MasterFrequency}, yielding~\cite{Barcelo2011,Tolosa2022}
\begin{equation}
    \omega_k^2 = k^2 - \frac{a''}{2a} + \left( \frac{a'}{2a} \right)^2 .
    \label{eq:BECFrequency}
\end{equation}
This corresponds to the situation where the coupling is minimal and therefore nonconformal, and second derivatives of the scale factor appear in the frequency. Regarding the density of produced particles, this BEC experiment corresponds to the scenario described by the non-conformal coupling curve $\xi=0$ in the leftmost panel of Fig.~\ref{fig:TotalDensity}.

\begin{figure}
    \centering
    \includegraphics[width=0.35\textwidth]{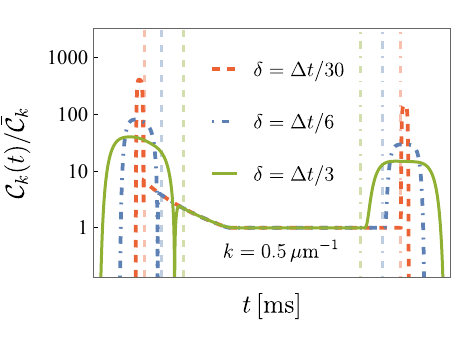}
    \caption{Function $\mathcal{C}_k$ for a typical analog gravity experiment in quasi-two-dimensional BECs for $k=0.5 \, \mu\text{m}^{-1}$ and transitions of different abruptness characterized by $\delta$. The values are normalized with respect to~$\bar{\mathcal{C}}_k \simeq 5 \times 10^{-11}$, assuming an intermediate region of duration~$\Delta t = 3 \, \text{ms}$. The BEC parameters correspond to the expansion linear in $t$ presented in Ref.~\cite{Viermann2022}.
    }
    \label{fig:CkBEC}
\end{figure}

We computed the function~$\mathcal{C}_k$ corresponding to the frequency~\eqref{eq:BECFrequency}, using the same functional form of the scale factor as in Eq.~\eqref{eq:DerivativeScale}, but replacing conformal time~$\eta$ with laboratory time $t$. In Fig. \ref{fig:CkBEC}, we replicate the laboratory conditions reported in Ref.~\cite{Viermann2022} for a scale factor linear in $t$ and observe that~$\mathcal{C}_k$ increases by several orders of magnitude during the `on' and `off' transitions compared to its lowest value during the expansion, $\bar{\mathcal{C}}_k$. This reinforces our earlier conclusion: The effects of abrupt transitions overshadow the contributions from the intermediate expansion process, effectively masking the dynamics we aim to analyze. One must, therefore, be aware of the role of transitions concerning particle production when performing such experiments, as the number of particles created stems from the `on' and `off' transitions rather than from the background time-dependence in the intermediate region.

Under laboratory conditions, it is more realistic to assume a nonzero initial occupation number~$n_k^0$, such as that of a thermal state, which results in stimulated particle production from the beginning. In this scenario, the expression for the expected particle number density~\eqref{eq:nk} is modified to~\mbox{$n_k = n_k^{0} + |\beta_k|^2 (1+2n_k^{0}$)} \cite{Tolosa2022,Viermann2022,Schmidt2024,Sparn2024}. This adjustment merely introduces an affine transformation. Therefore, the stimulated production of particles and antiparticles remains primarily dictated by the `on' and `off' transitions.

\section{Schwinger effect}
\label{sec:Schwinger}

In the Schwinger effect~\cite{Sauter1931,Schwinger1951}, a strong electromagnetic field excites a charged matter field in $(1+3)$-flat spacetime, resulting in particle creation. In this context, the role of the scale factor in the cosmological case is replaced by the electromagnetic potential~$A_\mu$. The Klein-Gordon equation for a scalar field~$\psi$ with mass~$m$ and charge~$q$ coupled to this potential is given by
\begin{equation}
\left[ (\partial_\mu + iqA_\mu)(\partial^\mu + iqA^\mu) - m^2 \right] \psi = 0.
\end{equation}
For a homogeneous, time-dependent electric field~$\textbf{E}(t)$, the temporal gauge \mbox{$A_{\mu}(t,\bfx) = (0, \textbf{A}(t))$} simplifies the equations of motion, making them homogeneous. In this gauge, the electric field is given by \mbox{$\textbf{E}(t) = -\dot{\textbf{A}}(t)$}. We consider that the direction of the electric field remains fixed over time. Without loss of generality, we align it along the $z$ axis.

Analogous to Eq.~\eqref{eq:ModeEquation} in the cosmological case, the Fourier modes~$\psi_{\bfk}$ obey harmonic oscillator equations of the form
\begin{equation}
    \frac{\text{d}^2}{\text{d}t^2} \psi_{\bfk}(t) + \Omega_{\bfk}^2(t) \psi_{\bfk}(t) = 0,
\end{equation}
where the time-dependent frequencies are given by
\begin{equation}
    \Omega_{\bfk}^2 = k^2 + 2q|\textbf{A}|\cos{\theta}k + q^2|\textbf{A}|^2 + m^2.
    \label{eq:SchwingerFrequency}
\end{equation}
The anisotropic nature of the system becomes evident in the form of~$\Omega_{\vk}$, as it depends on the angle~$\theta$ between the wave vector~$\vk$ and the direction of the vector potential~$\textbf{A}$ through a linear term in~$k$. 

\begin{figure}
    \centering
    \includegraphics[width=0.35\textwidth]{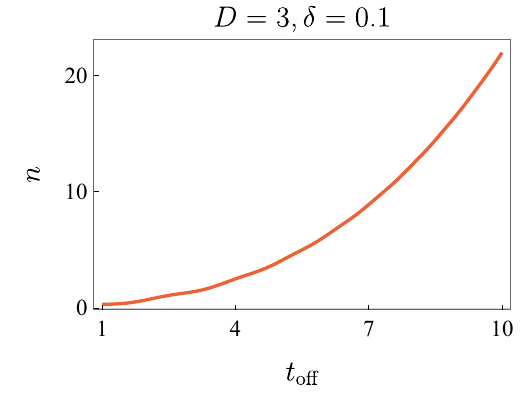}
    \caption{Total number density of produced particles $n$ as function of $t_{\text{off}}$ ($t_{\text{on}}=0$) in the Schwinger effect for electric potentials of the form~\eqref{eq:DerivativeScale}, with an intermediate electric strength equal to the Schwinger limit, $-\dot{A}_0 = m^2/q^2$, and fast switch-on and switch-off transitions (\mbox{$\delta=0.1$}). Time variables are expressed in units of $m^{-1}$, and the number density in units of $m^3$.}
    \label{fig:TotalDensitySchwinger}
\end{figure}

Since the frequency~$\Omega_{\bfk}$ is independent of any time derivatives of the potential, the analysis regarding its time variation yields conclusions similar to those in the case of a cosmologically conformally coupled scalar field. However, the intermediate regime has a more significant impact on the particle spectrum in the Schwinger effect than in the cosmological case. As shown in Fig.~\ref{fig:TotalDensitySchwinger}, the total number density of produced particles in the Schwinger effect continues to increase for large values of~$t_{\text{off}}$, in contrast to the cosmological case (Fig.~\ref{fig:TotalDensity}), where particle production eventually converges. This is a consequence of the linear dependence on~$k$ in the frequency~\eqref{eq:SchwingerFrequency} through the anisotropic term \mbox{$2 q |\textbf{A}| \cos\theta \, k$}. As the duration of the electric field increases, this term drives the excitation of higher-$k$ modes with $q \cos\theta < 0$, especially those aligned with the field direction, which provide the dominant contribution. This causes the intermediate regime to become increasingly dominant over abrupt transitions as the electric field remains switched on for a sufficiently long duration. This is consistent with Ref.~\cite{Adorno2018}, where it is demonstrated that, when the electric field remains on for sufficiently long durations, the dominant contribution to particle production comes from the intermediate regime rather than the switch-on and switch-off transitions. Nonetheless, even in this case---and even more so when the electric field is switched on for shorter durations---the effects of `on' and `off' transitions remain unavoidable.

\section{Conclusions}
\label{sec:conclusions}

In this work, we highlight the importance of carefully accounting for `on' and `off' transitions when interpreting quantum pair production due to the expansion of the Universe or a strong electromagnetic field. In experiments designed for simulating production within some time window, such regimes are inevitable---experiments have a beginning and an end---, and always influence particle production. Here, we have distinguished when these effects simply influence particle production without dominating it, and when they overwhelmingly dictate the outcome, necessitating a fundamental reinterpretation of the resulting spectra.

This issue is particularly critical in analog gravity experiments that simulate a nonconformally coupled field to an expanding FLRW universe, such as~\cite{Hung2013,Steinhauer2022,Viermann2022,Sparn2024}. From our analysis of the time dependence of the system's characteristic frequency, grounded in quantum kinetic arguments, we showed that transitions dominate particle production, effectively overshadowing the contributions from the intermediate dynamics. Therefore, one has to be careful when interpreting the outcomes of such experiments, as the main contribution to pair production does not come from the specific expansion during the intermediate phase that the setup is intended to simulate. In scenarios involving alternating periods of expansion and contraction~\cite{Schmidt2024,Sparn2024,Agullo:2024lry}, rapid changes in the scale factor significantly affect the particle spectrum, effectively acting as ‘on’ and ‘off’ transitions within the intermediate regime.

Interestingly, in situations where the test field is conformally coupled to the geometry, the contribution from the transitions no longer dominates, though it remains comparable to that of the intermediate regime of interest. 

The Schwinger effect presents a notably different scenario. While the `on' and `off' transitions still influence the outcome, in a similar way as in the conformally coupled cosmological case, their relative impact diminishes as the electric field remains switched on for longer times. In this regime, the intermediate period becomes increasingly dominant in determining the particle spectrum. This behavior stems from the intrinsic anisotropy of the electromagnetic background and stands in sharp contrast to the isotropic cosmological case, where extending the duration of the intermediate expansion has little effect on the spectrum, which remains dominated by the abrupt transitions. 

Studying how `on' and `off' transitions affect production is also of interest in the context of the early Universe, where particle production is typically computed from the onset of inflation until the expansion of spacetime slows down significantly, well into the reheating epoch. These periods behave as approximate `in' and `out' regions, where the mode frequency evolves very slowly, and between which `on' and `off' transitions occur.

Analog experiments inherently incorporate such transitions, and, if appropriately tuned, they could even simulate the \textit{actual} cosmological scenario---including the transition from inflation to reheating, where most particles are known to be produced~\cite{Ema2016,Markkanen2017b,Ema2018,Chung2019,Bastero2019,Yu2023,Cembranos2023}. However, it is crucial to abandon the idea of isolating the contribution of a specific intermediate region to pair production, as `on' and `off' transitions in such experiments remain unavoidable. As such, the measured particle spectra must be appropriately interpreted.

\begin{acknowledgments}
The authors would like to thank Mercedes Mart\'{i}n-Benito and Luis J. Garay for helpful discussions. This work is part of the R+D+I Projects No. PID2022-139841NB-I00 and No. PID2023-149018NB-C44, funded by MICIU/AEI/10.13039/501100011033 and by ERDF/EU. Additionally, A.P.L. acknowledges support through MICIU fellowship FPU20/05603, project No. PID2021- 127726NB-
I00 (MCIU/AEI/FEDER, UE), Grant IFT Centro de Excelencia Severo Ochoa No. CEX2020- 001007-S, funded by MCIN/AEI/10.13039/501100011033, the CAM/FEDER Project No. TEC-2024/COM 84 QUITEMAD-CM, and from the CSIC Research Platform on Quantum Technologies PTI-001. 
\end{acknowledgments}

\bibliography{Bibliography.bib}

\begin{thebibliography}{60}%
\makeatletter
\providecommand \@ifxundefined [1]{%
 \@ifx{#1\undefined}
}%
\providecommand \@ifnum [1]{%
 \ifnum #1\expandafter \@firstoftwo
 \else \expandafter \@secondoftwo
 \fi
}%
\providecommand \@ifx [1]{%
 \ifx #1\expandafter \@firstoftwo
 \else \expandafter \@secondoftwo
 \fi
}%
\providecommand \natexlab [1]{#1}%
\providecommand \enquote  [1]{``#1''}%
\providecommand \bibnamefont  [1]{#1}%
\providecommand \bibfnamefont [1]{#1}%
\providecommand \citenamefont [1]{#1}%
\providecommand \href@noop [0]{\@secondoftwo}%
\providecommand \href [0]{\begingroup \@sanitize@url \@href}%
\providecommand \@href[1]{\@@startlink{#1}\@@href}%
\providecommand \@@href[1]{\endgroup#1\@@endlink}%
\providecommand \@sanitize@url [0]{\catcode `\\12\catcode `\$12\catcode
  `\&12\catcode `\#12\catcode `\^12\catcode `\_12\catcode `\%12\relax}%
\providecommand \@@startlink[1]{}%
\providecommand \@@endlink[0]{}%
\providecommand \url  [0]{\begingroup\@sanitize@url \@url }%
\providecommand \@url [1]{\endgroup\@href {#1}{\urlprefix }}%
\providecommand \urlprefix  [0]{URL }%
\providecommand \Eprint [0]{\href }%
\providecommand \doibase [0]{https://doi.org/}%
\providecommand \selectlanguage [0]{\@gobble}%
\providecommand \bibinfo  [0]{\@secondoftwo}%
\providecommand \bibfield  [0]{\@secondoftwo}%
\providecommand \translation [1]{[#1]}%
\providecommand \BibitemOpen [0]{}%
\providecommand \bibitemStop [0]{}%
\providecommand \bibitemNoStop [0]{.\EOS\space}%
\providecommand \EOS [0]{\spacefactor3000\relax}%
\providecommand \BibitemShut  [1]{\csname bibitem#1\endcsname}%
\let\auto@bib@innerbib\@empty
\bibitem [{\citenamefont {Parker}(1968)}]{Parker1968}%
  \BibitemOpen
  \bibfield  {author} {\bibinfo {author} {\bibfnamefont {L.}~\bibnamefont
  {Parker}},\ }\bibfield  {title} {\bibinfo {title} {{Particle creation in
  expanding universes}},\ }\href {https://doi.org/10.1103/PhysRevLett.21.562}
  {\bibfield  {journal} {\bibinfo  {journal} {Phys. Rev. Lett.}\ }\textbf
  {\bibinfo {volume} {21}},\ \bibinfo {pages} {562} (\bibinfo {year}
  {1968})}\BibitemShut {NoStop}%
\bibitem [{\citenamefont {Parker}(1969)}]{Parker1969}%
  \BibitemOpen
  \bibfield  {author} {\bibinfo {author} {\bibfnamefont {L.}~\bibnamefont
  {Parker}},\ }\bibfield  {title} {\bibinfo {title} {{Quantized Fields and
  Particle Creation in Expanding Universes. I}},\ }\href
  {https://doi.org/10.1103/PhysRev.183.1057} {\bibfield  {journal} {\bibinfo
  {journal} {Phys. Rev.}\ }\textbf {\bibinfo {volume} {183}},\ \bibinfo {pages}
  {1057} (\bibinfo {year} {1969})}\BibitemShut {NoStop}%
\bibitem [{\citenamefont {Ford}(1987)}]{Ford1987}%
  \BibitemOpen
  \bibfield  {author} {\bibinfo {author} {\bibfnamefont {L.~H.}\ \bibnamefont
  {Ford}},\ }\bibfield  {title} {\bibinfo {title} {Gravitational particle
  creation and inflation},\ }\href {https://doi.org/10.1103/PhysRevD.35.2955}
  {\bibfield  {journal} {\bibinfo  {journal} {Phys. Rev. D}\ }\textbf {\bibinfo
  {volume} {35}},\ \bibinfo {pages} {2955} (\bibinfo {year}
  {1987})}\BibitemShut {NoStop}%
\bibitem [{\citenamefont {Hawking}(1974)}]{Hawking1974}%
  \BibitemOpen
  \bibfield  {author} {\bibinfo {author} {\bibfnamefont {S.~W.}\ \bibnamefont
  {Hawking}},\ }\bibfield  {title} {\bibinfo {title} {{Black hole
  explosions}},\ }\href {https://doi.org/10.1038/248030a0} {\bibfield
  {journal} {\bibinfo  {journal} {Nature}\ }\textbf {\bibinfo {volume} {248}},\
  \bibinfo {pages} {30} (\bibinfo {year} {1974})}\BibitemShut {NoStop}%
\bibitem [{\citenamefont {Hawking}(1975)}]{Hawking1975}%
  \BibitemOpen
  \bibfield  {author} {\bibinfo {author} {\bibfnamefont {S.~W.}\ \bibnamefont
  {Hawking}},\ }\bibfield  {title} {\bibinfo {title} {{Particle creation by
  black holes}},\ }\href {https://doi.org/10.1007/BF02345020} {\bibfield
  {journal} {\bibinfo  {journal} {Comm. Math. Phys.}\ }\textbf {\bibinfo
  {volume} {43}},\ \bibinfo {pages} {199} (\bibinfo {year} {1975})}\BibitemShut
  {NoStop}%
\bibitem [{\citenamefont {Sauter}(1931)}]{Sauter1931}%
  \BibitemOpen
  \bibfield  {author} {\bibinfo {author} {\bibfnamefont {F.}~\bibnamefont
  {Sauter}},\ }\bibfield  {title} {\bibinfo {title} {{\"Uber das Verhalten
  eines Elektrons im homogenen elektrischen Feld nach der relativistischen
  Theorie Diracs}},\ }\href {https://doi.org/10.1007/BF01339461} {\bibfield
  {journal} {\bibinfo  {journal} {Z. Phys.}\ }\textbf {\bibinfo {volume}
  {69}},\ \bibinfo {pages} {742} (\bibinfo {year} {1931})}\BibitemShut
  {NoStop}%
\bibitem [{\citenamefont {Heisenberg}\ and\ \citenamefont
  {Euler}(1936)}]{Heisenberg1936}%
  \BibitemOpen
  \bibfield  {author} {\bibinfo {author} {\bibfnamefont {W.}~\bibnamefont
  {Heisenberg}}\ and\ \bibinfo {author} {\bibfnamefont {H.}~\bibnamefont
  {Euler}},\ }\bibfield  {title} {\bibinfo {title} {{Folgerungen aus der
  Diracschen Theorie des Positrons}},\ }\href
  {https://doi.org/10.1007/BF01343663} {\bibfield  {journal} {\bibinfo
  {journal} {Z. Phys.}\ }\textbf {\bibinfo {volume} {98}},\ \bibinfo {pages}
  {714} (\bibinfo {year} {1936})}\BibitemShut {NoStop}%
\bibitem [{\citenamefont {Schwinger}(1951)}]{Schwinger1951}%
  \BibitemOpen
  \bibfield  {author} {\bibinfo {author} {\bibfnamefont {J.~S.}\ \bibnamefont
  {Schwinger}},\ }\bibfield  {title} {\bibinfo {title} {On gauge invariance and
  vacuum polarization},\ }\href {https://doi.org/10.1103/PhysRev.82.664}
  {\bibfield  {journal} {\bibinfo  {journal} {Phys. Rev.}\ }\textbf {\bibinfo
  {volume} {82}},\ \bibinfo {pages} {664} (\bibinfo {year} {1951})}\BibitemShut
  {NoStop}%
\bibitem [{\citenamefont {Unruh}(1981)}]{Unruh1981}%
  \BibitemOpen
  \bibfield  {author} {\bibinfo {author} {\bibfnamefont {W.~G.}\ \bibnamefont
  {Unruh}},\ }\bibfield  {title} {\bibinfo {title} {{Experimental Black-Hole
  Evaporation?}},\ }\href {https://doi.org/10.1103/PhysRevLett.46.1351}
  {\bibfield  {journal} {\bibinfo  {journal} {Phys. Rev. Lett.}\ }\textbf
  {\bibinfo {volume} {46}},\ \bibinfo {pages} {1351} (\bibinfo {year}
  {1981})}\BibitemShut {NoStop}%
\bibitem [{\citenamefont {Barceló}\ \emph {et~al.}(2011)\citenamefont
  {Barceló}, \citenamefont {Liberati},\ and\ \citenamefont
  {Visser}}]{Barcelo2011}%
  \BibitemOpen
  \bibfield  {author} {\bibinfo {author} {\bibfnamefont {C.}~\bibnamefont
  {Barceló}}, \bibinfo {author} {\bibfnamefont {S.}~\bibnamefont {Liberati}},\
  and\ \bibinfo {author} {\bibfnamefont {M.}~\bibnamefont {Visser}},\
  }\bibfield  {title} {\bibinfo {title} {{Analogue Gravity}},\ }\href
  {https://doi.org/10.12942/lrr-2011-3} {\bibfield  {journal} {\bibinfo
  {journal} {Living Rev. Relativ.}\ }\textbf {\bibinfo {volume} {14}},\
  \bibinfo {pages} {3} (\bibinfo {year} {2011})}\BibitemShut {NoStop}%
\bibitem [{\citenamefont {Jacquet}\ \emph {et~al.}(2020)\citenamefont
  {Jacquet}, \citenamefont {Weinfurtner},\ and\ \citenamefont
  {König}}]{Jacquet2020}%
  \BibitemOpen
  \bibfield  {author} {\bibinfo {author} {\bibfnamefont {M.~J.}\ \bibnamefont
  {Jacquet}}, \bibinfo {author} {\bibfnamefont {S.}~\bibnamefont
  {Weinfurtner}},\ and\ \bibinfo {author} {\bibfnamefont {F.}~\bibnamefont
  {König}},\ }\bibfield  {title} {\bibinfo {title} {{The next generation of
  analogue gravity experiments}},\ }\href
  {https://doi.org/10.1098/rsta.2019.0239} {\bibfield  {journal} {\bibinfo
  {journal} {Philos. Trans. Royal Soc. A}\ }\textbf {\bibinfo {volume} {378}},\
  \bibinfo {pages} {20190239} (\bibinfo {year} {2020})}\BibitemShut {NoStop}%
\bibitem [{\citenamefont {Almeida}\ and\ \citenamefont
  {Jacquet}(2023)}]{Almeida2023}%
  \BibitemOpen
  \bibfield  {author} {\bibinfo {author} {\bibfnamefont {C.~R.}\ \bibnamefont
  {Almeida}}\ and\ \bibinfo {author} {\bibfnamefont {M.~J.}\ \bibnamefont
  {Jacquet}},\ }\bibfield  {title} {\bibinfo {title} {{Analogue gravity and the
  Hawking effect: historical perspective and literature review}},\ }\href
  {https://doi.org/10.1140/epjh/s13129-023-00063-2} {\bibfield  {journal}
  {\bibinfo  {journal} {Eur. Phys. J. H}\ }\textbf {\bibinfo {volume} {48}},\
  \bibinfo {pages} {15} (\bibinfo {year} {2023})}\BibitemShut {NoStop}%
\bibitem [{\citenamefont {Philbin}\ \emph {et~al.}(2008)\citenamefont
  {Philbin}, \citenamefont {Kuklewicz}, \citenamefont {Robertson},
  \citenamefont {Hill}, \citenamefont {König},\ and\ \citenamefont
  {Leonhardt}}]{Philbin2008}%
  \BibitemOpen
  \bibfield  {author} {\bibinfo {author} {\bibfnamefont {T.~G.}\ \bibnamefont
  {Philbin}}, \bibinfo {author} {\bibfnamefont {C.}~\bibnamefont {Kuklewicz}},
  \bibinfo {author} {\bibfnamefont {S.}~\bibnamefont {Robertson}}, \bibinfo
  {author} {\bibfnamefont {S.}~\bibnamefont {Hill}}, \bibinfo {author}
  {\bibfnamefont {F.}~\bibnamefont {König}},\ and\ \bibinfo {author}
  {\bibfnamefont {U.}~\bibnamefont {Leonhardt}},\ }\bibfield  {title} {\bibinfo
  {title} {{Fiber-Optical Analog of the Event Horizon}},\ }\href
  {https://doi.org/10.1126/science.1153625} {\bibfield  {journal} {\bibinfo
  {journal} {Science}\ }\textbf {\bibinfo {volume} {319}},\ \bibinfo {pages}
  {1367} (\bibinfo {year} {2008})}\BibitemShut {NoStop}%
\bibitem [{\citenamefont {Weinfurtner}\ \emph {et~al.}(2011)\citenamefont
  {Weinfurtner}, \citenamefont {Tedford}, \citenamefont {Penrice},
  \citenamefont {Unruh},\ and\ \citenamefont {Lawrence}}]{Weinfurtner2011}%
  \BibitemOpen
  \bibfield  {author} {\bibinfo {author} {\bibfnamefont {S.}~\bibnamefont
  {Weinfurtner}}, \bibinfo {author} {\bibfnamefont {E.~W.}\ \bibnamefont
  {Tedford}}, \bibinfo {author} {\bibfnamefont {M.~C.~J.}\ \bibnamefont
  {Penrice}}, \bibinfo {author} {\bibfnamefont {W.~G.}\ \bibnamefont {Unruh}},\
  and\ \bibinfo {author} {\bibfnamefont {G.~A.}\ \bibnamefont {Lawrence}},\
  }\bibfield  {title} {\bibinfo {title} {{Measurement of Stimulated Hawking
  Emission in an Analogue System}},\ }\href
  {https://doi.org/10.1103/PhysRevLett.106.021302} {\bibfield  {journal}
  {\bibinfo  {journal} {Phys. Rev. Lett.}\ }\textbf {\bibinfo {volume} {106}},\
  \bibinfo {pages} {021302} (\bibinfo {year} {2011})}\BibitemShut {NoStop}%
\bibitem [{\citenamefont {Euv\'e}\ \emph {et~al.}(2016)\citenamefont {Euv\'e},
  \citenamefont {Michel}, \citenamefont {Parentani}, \citenamefont {Philbin},\
  and\ \citenamefont {Rousseaux}}]{Euve2016}%
  \BibitemOpen
  \bibfield  {author} {\bibinfo {author} {\bibfnamefont {L.-P.}\ \bibnamefont
  {Euv\'e}}, \bibinfo {author} {\bibfnamefont {F.}~\bibnamefont {Michel}},
  \bibinfo {author} {\bibfnamefont {R.}~\bibnamefont {Parentani}}, \bibinfo
  {author} {\bibfnamefont {T.~G.}\ \bibnamefont {Philbin}},\ and\ \bibinfo
  {author} {\bibfnamefont {G.}~\bibnamefont {Rousseaux}},\ }\bibfield  {title}
  {\bibinfo {title} {Observation of noise correlated by the {Hawking} effect in
  a water tank},\ }\href {https://doi.org/10.1103/PhysRevLett.117.121301}
  {\bibfield  {journal} {\bibinfo  {journal} {Phys. Rev. Lett.}\ }\textbf
  {\bibinfo {volume} {117}},\ \bibinfo {pages} {121301} (\bibinfo {year}
  {2016})}\BibitemShut {NoStop}%
\bibitem [{\citenamefont {Muñoz~de Nova}\ \emph {et~al.}(2019)\citenamefont
  {Muñoz~de Nova}, \citenamefont {Golubkov}, \citenamefont {Kolobov},\ and\
  \citenamefont {Steinhauer}}]{MunozDeNova2019}%
  \BibitemOpen
  \bibfield  {author} {\bibinfo {author} {\bibfnamefont {J.~R.}\ \bibnamefont
  {Muñoz~de Nova}}, \bibinfo {author} {\bibfnamefont {K.}~\bibnamefont
  {Golubkov}}, \bibinfo {author} {\bibfnamefont {V.~I.}\ \bibnamefont
  {Kolobov}},\ and\ \bibinfo {author} {\bibfnamefont {J.}~\bibnamefont
  {Steinhauer}},\ }\bibfield  {title} {\bibinfo {title} {{Observation of
  thermal Hawking radiation and its temperature in an analogue black hole}},\
  }\href {https://doi.org/10.1038/s41586-019-1241-0} {\bibfield  {journal}
  {\bibinfo  {journal} {Nature}\ }\textbf {\bibinfo {volume} {569}},\ \bibinfo
  {pages} {688} (\bibinfo {year} {2019})}\BibitemShut {NoStop}%
\bibitem [{\citenamefont {Drori}\ \emph {et~al.}(2019)\citenamefont {Drori},
  \citenamefont {Rosenberg}, \citenamefont {Bermudez}, \citenamefont
  {Silberberg},\ and\ \citenamefont {Leonhardt}}]{Drori2019}%
  \BibitemOpen
  \bibfield  {author} {\bibinfo {author} {\bibfnamefont {J.}~\bibnamefont
  {Drori}}, \bibinfo {author} {\bibfnamefont {Y.}~\bibnamefont {Rosenberg}},
  \bibinfo {author} {\bibfnamefont {D.}~\bibnamefont {Bermudez}}, \bibinfo
  {author} {\bibfnamefont {Y.}~\bibnamefont {Silberberg}},\ and\ \bibinfo
  {author} {\bibfnamefont {U.}~\bibnamefont {Leonhardt}},\ }\bibfield  {title}
  {\bibinfo {title} {Observation of {Stimulated} {Hawking} {Radiation} in an
  {Optical} {Analogue}},\ }\href
  {https://doi.org/10.1103/PhysRevLett.122.010404} {\bibfield  {journal}
  {\bibinfo  {journal} {Phys. Rev. Lett.}\ }\textbf {\bibinfo {volume} {122}},\
  \bibinfo {pages} {010404} (\bibinfo {year} {2019})}\BibitemShut {NoStop}%
\bibitem [{\citenamefont {Shi}\ \emph {et~al.}(2023)\citenamefont {Shi},
  \citenamefont {Yang}, \citenamefont {Xiang}, \citenamefont {Ge},
  \citenamefont {Li}, \citenamefont {Wang}, \citenamefont {Huang},
  \citenamefont {Tian}, \citenamefont {Song}, \citenamefont {Zheng},
  \citenamefont {Xu}, \citenamefont {Cai},\ and\ \citenamefont
  {Fan}}]{Shi2023}%
  \BibitemOpen
  \bibfield  {author} {\bibinfo {author} {\bibfnamefont {Y.-H.}\ \bibnamefont
  {Shi}}, \bibinfo {author} {\bibfnamefont {R.-Q.}\ \bibnamefont {Yang}},
  \bibinfo {author} {\bibfnamefont {Z.}~\bibnamefont {Xiang}}, \bibinfo
  {author} {\bibfnamefont {Z.-Y.}\ \bibnamefont {Ge}}, \bibinfo {author}
  {\bibfnamefont {H.}~\bibnamefont {Li}}, \bibinfo {author} {\bibfnamefont
  {Y.-Y.}\ \bibnamefont {Wang}}, \bibinfo {author} {\bibfnamefont
  {K.}~\bibnamefont {Huang}}, \bibinfo {author} {\bibfnamefont
  {Y.}~\bibnamefont {Tian}}, \bibinfo {author} {\bibfnamefont {X.}~\bibnamefont
  {Song}}, \bibinfo {author} {\bibfnamefont {D.}~\bibnamefont {Zheng}},
  \bibinfo {author} {\bibfnamefont {K.}~\bibnamefont {Xu}}, \bibinfo {author}
  {\bibfnamefont {R.-G.}\ \bibnamefont {Cai}},\ and\ \bibinfo {author}
  {\bibfnamefont {H.}~\bibnamefont {Fan}},\ }\bibfield  {title} {\bibinfo
  {title} {Quantum simulation of {Hawking} radiation and curved spacetime with
  a superconducting on-chip black hole},\ }\href
  {https://doi.org/10.1038/s41467-023-39064-6} {\bibfield  {journal} {\bibinfo
  {journal} {Nat. Commun.}\ }\textbf {\bibinfo {volume} {14}},\ \bibinfo
  {pages} {3263} (\bibinfo {year} {2023})}\BibitemShut {NoStop}%
\bibitem [{\citenamefont {Torres}\ \emph {et~al.}(2017)\citenamefont {Torres},
  \citenamefont {Patrick}, \citenamefont {Coutant}, \citenamefont {Richartz},
  \citenamefont {Tedford},\ and\ \citenamefont {Weinfurtner}}]{Torres2017}%
  \BibitemOpen
  \bibfield  {author} {\bibinfo {author} {\bibfnamefont {T.}~\bibnamefont
  {Torres}}, \bibinfo {author} {\bibfnamefont {S.}~\bibnamefont {Patrick}},
  \bibinfo {author} {\bibfnamefont {A.}~\bibnamefont {Coutant}}, \bibinfo
  {author} {\bibfnamefont {M.}~\bibnamefont {Richartz}}, \bibinfo {author}
  {\bibfnamefont {E.~W.}\ \bibnamefont {Tedford}},\ and\ \bibinfo {author}
  {\bibfnamefont {S.}~\bibnamefont {Weinfurtner}},\ }\bibfield  {title}
  {\bibinfo {title} {Rotational superradiant scattering in a vortex flow},\
  }\href {https://doi.org/10.1038/nphys4151} {\bibfield  {journal} {\bibinfo
  {journal} {Nat. Phys.}\ }\textbf {\bibinfo {volume} {13}},\ \bibinfo {pages}
  {833} (\bibinfo {year} {2017})}\BibitemShut {NoStop}%
\bibitem [{\citenamefont {Eckel}\ \emph {et~al.}(2018)\citenamefont {Eckel},
  \citenamefont {Kumar}, \citenamefont {Jacobson}, \citenamefont {Spielman},\
  and\ \citenamefont {Campbell}}]{Eckel2018}%
  \BibitemOpen
  \bibfield  {author} {\bibinfo {author} {\bibfnamefont {S.}~\bibnamefont
  {Eckel}}, \bibinfo {author} {\bibfnamefont {A.}~\bibnamefont {Kumar}},
  \bibinfo {author} {\bibfnamefont {T.}~\bibnamefont {Jacobson}}, \bibinfo
  {author} {\bibfnamefont {I.~B.}\ \bibnamefont {Spielman}},\ and\ \bibinfo
  {author} {\bibfnamefont {G.~K.}\ \bibnamefont {Campbell}},\ }\bibfield
  {title} {\bibinfo {title} {{A Rapidly Expanding Bose-Einstein Condensate: An
  Expanding Universe in the Lab}},\ }\href
  {https://doi.org/10.1103/PhysRevX.8.021021} {\bibfield  {journal} {\bibinfo
  {journal} {Phys. Rev. X}\ }\textbf {\bibinfo {volume} {8}},\ \bibinfo {pages}
  {021021} (\bibinfo {year} {2018})}\BibitemShut {NoStop}%
\bibitem [{\citenamefont {Wittemer}\ \emph {et~al.}(2019)\citenamefont
  {Wittemer}, \citenamefont {Hakelberg}, \citenamefont {Kiefer}, \citenamefont
  {Schr\"oder}, \citenamefont {Fey}, \citenamefont {Sch\"utzhold},
  \citenamefont {Warring},\ and\ \citenamefont {Schaetz}}]{Wittemer2019}%
  \BibitemOpen
  \bibfield  {author} {\bibinfo {author} {\bibfnamefont {M.}~\bibnamefont
  {Wittemer}}, \bibinfo {author} {\bibfnamefont {F.}~\bibnamefont {Hakelberg}},
  \bibinfo {author} {\bibfnamefont {P.}~\bibnamefont {Kiefer}}, \bibinfo
  {author} {\bibfnamefont {J.-P.}\ \bibnamefont {Schr\"oder}}, \bibinfo
  {author} {\bibfnamefont {C.}~\bibnamefont {Fey}}, \bibinfo {author}
  {\bibfnamefont {R.}~\bibnamefont {Sch\"utzhold}}, \bibinfo {author}
  {\bibfnamefont {U.}~\bibnamefont {Warring}},\ and\ \bibinfo {author}
  {\bibfnamefont {T.}~\bibnamefont {Schaetz}},\ }\bibfield  {title} {\bibinfo
  {title} {{Phonon Pair Creation by Inflating Quantum Fluctuations in an Ion
  Trap}},\ }\href {https://doi.org/10.1103/PhysRevLett.123.180502} {\bibfield
  {journal} {\bibinfo  {journal} {Phys. Rev. Lett.}\ }\textbf {\bibinfo
  {volume} {123}},\ \bibinfo {pages} {180502} (\bibinfo {year}
  {2019})}\BibitemShut {NoStop}%
\bibitem [{\citenamefont {Banik}\ \emph {et~al.}(2022)\citenamefont {Banik},
  \citenamefont {Galan}, \citenamefont {Sosa-Martinez}, \citenamefont
  {Anderson}, \citenamefont {Eckel}, \citenamefont {Spielman},\ and\
  \citenamefont {Campbell}}]{Banik2021}%
  \BibitemOpen
  \bibfield  {author} {\bibinfo {author} {\bibfnamefont {S.}~\bibnamefont
  {Banik}}, \bibinfo {author} {\bibfnamefont {M.~G.}\ \bibnamefont {Galan}},
  \bibinfo {author} {\bibfnamefont {H.}~\bibnamefont {Sosa-Martinez}}, \bibinfo
  {author} {\bibfnamefont {M.}~\bibnamefont {Anderson}}, \bibinfo {author}
  {\bibfnamefont {S.}~\bibnamefont {Eckel}}, \bibinfo {author} {\bibfnamefont
  {I.~B.}\ \bibnamefont {Spielman}},\ and\ \bibinfo {author} {\bibfnamefont
  {G.~K.}\ \bibnamefont {Campbell}},\ }\bibfield  {title} {\bibinfo {title}
  {{Accurate Determination of Hubble Attenuation and Amplification in Expanding
  and Contracting Cold-Atom Universes}},\ }\href
  {https://doi.org/10.1103/PhysRevLett.128.090401} {\bibfield  {journal}
  {\bibinfo  {journal} {Phys. Rev. Lett.}\ }\textbf {\bibinfo {volume} {128}},\
  \bibinfo {pages} {090401} (\bibinfo {year} {2022})}\BibitemShut {NoStop}%
\bibitem [{\citenamefont {Giacomelli}\ and\ \citenamefont
  {Carusotto}(2021)}]{Giacomelli2021}%
  \BibitemOpen
  \bibfield  {author} {\bibinfo {author} {\bibfnamefont {L.}~\bibnamefont
  {Giacomelli}}\ and\ \bibinfo {author} {\bibfnamefont {I.}~\bibnamefont
  {Carusotto}},\ }\bibfield  {title} {\bibinfo {title} {{Understanding
  superradiant phenomena with synthetic vector potentials in atomic
  Bose-Einstein condensates}},\ }\href
  {http://dx.doi.org/10.1103/PhysRevA.103.043309} {\bibfield  {journal}
  {\bibinfo  {journal} {Phys. Rev. A}\ }\textbf {\bibinfo {volume} {103}},\
  \bibinfo {pages} {043309} (\bibinfo {year} {2021})}\BibitemShut {NoStop}%
\bibitem [{\citenamefont {Braidotti}\ \emph {et~al.}(2022)\citenamefont
  {Braidotti}, \citenamefont {Prizia}, \citenamefont {Maitland}, \citenamefont
  {Marino}, \citenamefont {Prain}, \citenamefont {Starshynov}, \citenamefont
  {Westerberg}, \citenamefont {Wright},\ and\ \citenamefont
  {Faccio}}]{Braidotti2022}%
  \BibitemOpen
  \bibfield  {author} {\bibinfo {author} {\bibfnamefont {M.-C.}\ \bibnamefont
  {Braidotti}}, \bibinfo {author} {\bibfnamefont {R.}~\bibnamefont {Prizia}},
  \bibinfo {author} {\bibfnamefont {C.}~\bibnamefont {Maitland}}, \bibinfo
  {author} {\bibfnamefont {F.}~\bibnamefont {Marino}}, \bibinfo {author}
  {\bibfnamefont {A.}~\bibnamefont {Prain}}, \bibinfo {author} {\bibfnamefont
  {I.}~\bibnamefont {Starshynov}}, \bibinfo {author} {\bibfnamefont
  {N.}~\bibnamefont {Westerberg}}, \bibinfo {author} {\bibfnamefont {E.~M.}\
  \bibnamefont {Wright}},\ and\ \bibinfo {author} {\bibfnamefont
  {D.}~\bibnamefont {Faccio}},\ }\bibfield  {title} {\bibinfo {title}
  {Measurement of {Penrose} {Superradiance} in a {Photon} {Superfluid}},\
  }\href {https://doi.org/10.1103/PhysRevLett.128.013901} {\bibfield  {journal}
  {\bibinfo  {journal} {Phys. Rev. Lett.}\ }\textbf {\bibinfo {volume} {128}},\
  \bibinfo {pages} {013901} (\bibinfo {year} {2022})}\BibitemShut {NoStop}%
\bibitem [{\citenamefont {Steinhauer}\ \emph {et~al.}(2022)\citenamefont
  {Steinhauer}, \citenamefont {Abuzarli}, \citenamefont {Aladjidi},
  \citenamefont {Bienaim{\'{e}}}, \citenamefont {Piekarski}, \citenamefont
  {Liu}, \citenamefont {Giacobino}, \citenamefont {Bramati},\ and\
  \citenamefont {Glorieux}}]{Steinhauer2022}%
  \BibitemOpen
  \bibfield  {author} {\bibinfo {author} {\bibfnamefont {J.}~\bibnamefont
  {Steinhauer}}, \bibinfo {author} {\bibfnamefont {M.}~\bibnamefont
  {Abuzarli}}, \bibinfo {author} {\bibfnamefont {T.}~\bibnamefont {Aladjidi}},
  \bibinfo {author} {\bibfnamefont {T.}~\bibnamefont {Bienaim{\'{e}}}},
  \bibinfo {author} {\bibfnamefont {C.}~\bibnamefont {Piekarski}}, \bibinfo
  {author} {\bibfnamefont {W.}~\bibnamefont {Liu}}, \bibinfo {author}
  {\bibfnamefont {E.}~\bibnamefont {Giacobino}}, \bibinfo {author}
  {\bibfnamefont {A.}~\bibnamefont {Bramati}},\ and\ \bibinfo {author}
  {\bibfnamefont {Q.}~\bibnamefont {Glorieux}},\ }\bibfield  {title} {\bibinfo
  {title} {Analogue cosmological particle creation in an ultracold quantum
  fluid of light},\ }\href {https://doi.org/10.1038/s41467-022-30603-1}
  {\bibfield  {journal} {\bibinfo  {journal} {Nat. Commun.}\ }\textbf {\bibinfo
  {volume} {13}},\ \bibinfo {pages} {2890} (\bibinfo {year}
  {2022})}\BibitemShut {NoStop}%
\bibitem [{\citenamefont {Jacquet}\ \emph {et~al.}(2022)\citenamefont
  {Jacquet}, \citenamefont {Joly}, \citenamefont {Claude}, \citenamefont
  {Giacomelli}, \citenamefont {Glorieux}, \citenamefont {Bramati},
  \citenamefont {Carusotto},\ and\ \citenamefont {Giacobino}}]{Jacquet2022}%
  \BibitemOpen
  \bibfield  {author} {\bibinfo {author} {\bibfnamefont {M.}~\bibnamefont
  {Jacquet}}, \bibinfo {author} {\bibfnamefont {M.}~\bibnamefont {Joly}},
  \bibinfo {author} {\bibfnamefont {F.}~\bibnamefont {Claude}}, \bibinfo
  {author} {\bibfnamefont {L.}~\bibnamefont {Giacomelli}}, \bibinfo {author}
  {\bibfnamefont {Q.}~\bibnamefont {Glorieux}}, \bibinfo {author}
  {\bibfnamefont {A.}~\bibnamefont {Bramati}}, \bibinfo {author} {\bibfnamefont
  {I.}~\bibnamefont {Carusotto}},\ and\ \bibinfo {author} {\bibfnamefont
  {E.}~\bibnamefont {Giacobino}},\ }\bibfield  {title} {\bibinfo {title}
  {{Analogue quantum simulation of the Hawking effect in a polariton
  superfluid}},\ }\href {http://dx.doi.org/10.1140/epjd/s10053-022-00477-5}
  {\bibfield  {journal} {\bibinfo  {journal} {Eur. Phys. J. D}\ }\textbf
  {\bibinfo {volume} {76}},\ \bibinfo {pages} {152} (\bibinfo {year}
  {2022})}\BibitemShut {NoStop}%
\bibitem [{\citenamefont {Viermann}\ \emph {et~al.}(2022)\citenamefont
  {Viermann}, \citenamefont {Sparn}, \citenamefont {Liebster}, \citenamefont
  {Hans}, \citenamefont {Kath}, \citenamefont {Parra-L\'opez}, \citenamefont
  {Tolosa-Sime\'on}, \citenamefont {S\'anchez-Kuntz}, \citenamefont {Haas},
  \citenamefont {Strobel}, \citenamefont {Floerchinger},\ and\ \citenamefont
  {Oberthaler}}]{Viermann2022}%
  \BibitemOpen
  \bibfield  {author} {\bibinfo {author} {\bibfnamefont {C.}~\bibnamefont
  {Viermann}}, \bibinfo {author} {\bibfnamefont {M.}~\bibnamefont {Sparn}},
  \bibinfo {author} {\bibfnamefont {N.}~\bibnamefont {Liebster}}, \bibinfo
  {author} {\bibfnamefont {M.}~\bibnamefont {Hans}}, \bibinfo {author}
  {\bibfnamefont {E.}~\bibnamefont {Kath}}, \bibinfo {author} {\bibfnamefont
  {A.}~\bibnamefont {Parra-L\'opez}}, \bibinfo {author} {\bibfnamefont
  {M.}~\bibnamefont {Tolosa-Sime\'on}}, \bibinfo {author} {\bibfnamefont
  {N.}~\bibnamefont {S\'anchez-Kuntz}}, \bibinfo {author} {\bibfnamefont
  {T.}~\bibnamefont {Haas}}, \bibinfo {author} {\bibfnamefont {H.}~\bibnamefont
  {Strobel}}, \bibinfo {author} {\bibfnamefont {S.}~\bibnamefont
  {Floerchinger}},\ and\ \bibinfo {author} {\bibfnamefont {M.~K.}\ \bibnamefont
  {Oberthaler}},\ }\bibfield  {title} {\bibinfo {title} {Quantum field
  simulator for dynamics in curved spacetime},\ }\href
  {https://doi.org/10.1038/s41586-022-05313-9} {\bibfield  {journal} {\bibinfo
  {journal} {Nature}\ }\textbf {\bibinfo {volume} {611}},\ \bibinfo {pages}
  {260} (\bibinfo {year} {2022})}\BibitemShut {NoStop}%
\bibitem [{\citenamefont {Sparn}\ \emph {et~al.}(2024)\citenamefont {Sparn},
  \citenamefont {Kath}, \citenamefont {Liebster}, \citenamefont {Duchene},
  \citenamefont {Schmidt}, \citenamefont {Tolosa-Sime\'on}, \citenamefont
  {Parra-L\'opez}, \citenamefont {Floerchinger}, \citenamefont {Strobel},\ and\
  \citenamefont {Oberthaler}}]{Sparn2024}%
  \BibitemOpen
  \bibfield  {author} {\bibinfo {author} {\bibfnamefont {M.}~\bibnamefont
  {Sparn}}, \bibinfo {author} {\bibfnamefont {E.}~\bibnamefont {Kath}},
  \bibinfo {author} {\bibfnamefont {N.}~\bibnamefont {Liebster}}, \bibinfo
  {author} {\bibfnamefont {J.}~\bibnamefont {Duchene}}, \bibinfo {author}
  {\bibfnamefont {C.~F.}\ \bibnamefont {Schmidt}}, \bibinfo {author}
  {\bibfnamefont {M.}~\bibnamefont {Tolosa-Sime\'on}}, \bibinfo {author}
  {\bibfnamefont {A.}~\bibnamefont {Parra-L\'opez}}, \bibinfo {author}
  {\bibfnamefont {S.}~\bibnamefont {Floerchinger}}, \bibinfo {author}
  {\bibfnamefont {H.}~\bibnamefont {Strobel}},\ and\ \bibinfo {author}
  {\bibfnamefont {M.~K.}\ \bibnamefont {Oberthaler}},\ }\bibfield  {title}
  {\bibinfo {title} {Experimental particle production in time-dependent
  spacetimes: A one-dimensional scattering problem},\ }\href
  {https://doi.org/10.1103/PhysRevLett.133.260201} {\bibfield  {journal}
  {\bibinfo  {journal} {Phys. Rev. Lett.}\ }\textbf {\bibinfo {volume} {133}},\
  \bibinfo {pages} {260201} (\bibinfo {year} {2024})}\BibitemShut {NoStop}%
\bibitem [{\citenamefont {Sch{\"u}tzhold}(2009)}]{Schuetzhold2009}%
  \BibitemOpen
  \bibfield  {author} {\bibinfo {author} {\bibfnamefont {R.}~\bibnamefont
  {Sch{\"u}tzhold}},\ }\bibfield  {title} {\bibinfo {title} {{Recreating
  Fundamental Effects in the Laboratory?}},\ }\href
  {https://doi.org/10.1166/asl.2009.1020} {\bibfield  {journal} {\bibinfo
  {journal} {Adv. Sci. Lett.}\ }\textbf {\bibinfo {volume} {2}},\ \bibinfo
  {pages} {121} (\bibinfo {year} {2009})}\BibitemShut {NoStop}%
\bibitem [{\citenamefont {Bulanov}\ \emph {et~al.}(2010)\citenamefont
  {Bulanov}, \citenamefont {Mur}, \citenamefont {Narozhny}, \citenamefont
  {Nees},\ and\ \citenamefont {Popov}}]{Bulanov2010}%
  \BibitemOpen
  \bibfield  {author} {\bibinfo {author} {\bibfnamefont {S.~S.}\ \bibnamefont
  {Bulanov}}, \bibinfo {author} {\bibfnamefont {V.~D.}\ \bibnamefont {Mur}},
  \bibinfo {author} {\bibfnamefont {N.~B.}\ \bibnamefont {Narozhny}}, \bibinfo
  {author} {\bibfnamefont {J.}~\bibnamefont {Nees}},\ and\ \bibinfo {author}
  {\bibfnamefont {V.~S.}\ \bibnamefont {Popov}},\ }\bibfield  {title} {\bibinfo
  {title} {{Multiple Colliding Electromagnetic Pulses: A Way to Lower the
  Threshold of ${e}^{+}{e}^{\ensuremath{-}}$ Pair Production from Vacuum}},\
  }\href {https://doi.org/10.1103/PhysRevLett.104.220404} {\bibfield  {journal}
  {\bibinfo  {journal} {Phys. Rev. Lett.}\ }\textbf {\bibinfo {volume} {104}},\
  \bibinfo {pages} {220404} (\bibinfo {year} {2010})}\BibitemShut {NoStop}%
\bibitem [{\citenamefont {Aleksandrov}\ \emph {et~al.}(2022)\citenamefont
  {Aleksandrov}, \citenamefont {Sevostyanov},\ and\ \citenamefont
  {Shabaev}}]{Aleksandrov2022}%
  \BibitemOpen
  \bibfield  {author} {\bibinfo {author} {\bibfnamefont {I.~A.}\ \bibnamefont
  {Aleksandrov}}, \bibinfo {author} {\bibfnamefont {D.~G.}\ \bibnamefont
  {Sevostyanov}},\ and\ \bibinfo {author} {\bibfnamefont {V.~M.}\ \bibnamefont
  {Shabaev}},\ }\bibfield  {title} {\bibinfo {title} {{Particle production in
  strong electromagnetic fields and local approximations}},\ }\href
  {https://doi.org/10.3390/sym14112444} {\bibfield  {journal} {\bibinfo
  {journal} {Symmetry}\ }\textbf {\bibinfo {volume} {14}},\ \bibinfo {pages}
  {2444} (\bibinfo {year} {2022})}\BibitemShut {NoStop}%
\bibitem [{\citenamefont {Aleksandrov}\ \emph {et~al.}(2025)\citenamefont
  {Aleksandrov}, \citenamefont {Sevostyanov},\ and\ \citenamefont
  {Shabaev}}]{Aleksandrov2025}%
  \BibitemOpen
  \bibfield  {author} {\bibinfo {author} {\bibfnamefont {I.~A.}\ \bibnamefont
  {Aleksandrov}}, \bibinfo {author} {\bibfnamefont {D.~G.}\ \bibnamefont
  {Sevostyanov}},\ and\ \bibinfo {author} {\bibfnamefont {V.~M.}\ \bibnamefont
  {Shabaev}},\ }\bibfield  {title} {\bibinfo {title} {Schwinger particle
  production: rapid switch off of the external field versus dynamical
  assistance},\ }\href {https://doi.org/10/g9gkfv} {\bibfield  {journal}
  {\bibinfo  {journal} {Phys. Rev. D}\ }\textbf {\bibinfo {volume} {111}},\
  \bibinfo {pages} {016010} (\bibinfo {year} {2025})}\BibitemShut {NoStop}%
\bibitem [{\citenamefont {Ilderton}(2022)}]{Ilderton2022}%
  \BibitemOpen
  \bibfield  {author} {\bibinfo {author} {\bibfnamefont {A.}~\bibnamefont
  {Ilderton}},\ }\bibfield  {title} {\bibinfo {title} {{Physics of adiabatic
  particle number in the Schwinger effect}},\ }\href
  {https://doi.org/10.1103/PhysRevD.105.016021} {\bibfield  {journal} {\bibinfo
   {journal} {Phys. Rev. D}\ }\textbf {\bibinfo {volume} {105}},\ \bibinfo
  {pages} {016021} (\bibinfo {year} {2022})}\BibitemShut {NoStop}%
\bibitem [{\citenamefont {Friedman}(1922)}]{Friedman1922}%
  \BibitemOpen
  \bibfield  {author} {\bibinfo {author} {\bibfnamefont {A.}~\bibnamefont
  {Friedman}},\ }\bibfield  {title} {\bibinfo {title} {{Über die Krümmung des
  Raumes}},\ }\href {https://doi.org/10.1007/BF01332580} {\bibfield  {journal}
  {\bibinfo  {journal} {Z. Phys.}\ }\textbf {\bibinfo {volume} {10}},\ \bibinfo
  {pages} {377} (\bibinfo {year} {1922})}\BibitemShut {NoStop}%
\bibitem [{\citenamefont {Friedman}(1924)}]{Friedman1924}%
  \BibitemOpen
  \bibfield  {author} {\bibinfo {author} {\bibfnamefont {A.}~\bibnamefont
  {Friedman}},\ }\bibfield  {title} {\bibinfo {title} {{Über die Möglichkeit
  einer Welt mit konstanter negativer Krümmung des Raumes}},\ }\href
  {https://doi.org/10.1007/BF01328280} {\bibfield  {journal} {\bibinfo
  {journal} {Z. Phys.}\ }\textbf {\bibinfo {volume} {21}},\ \bibinfo {pages}
  {326} (\bibinfo {year} {1924})}\BibitemShut {NoStop}%
\bibitem [{\citenamefont {Lemaître}(1931)}]{Lemaitre1931}%
  \BibitemOpen
  \bibfield  {author} {\bibinfo {author} {\bibfnamefont {A.~G.}\ \bibnamefont
  {Lemaître}},\ }\bibfield  {title} {\bibinfo {title} {{A Homogeneous Universe
  of Constant Mass and Increasing Radius accounting for the Radial Velocity of
  Extra-galactic Nebulæ}},\ }\href {https://doi.org/10.1093/mnras/91.5.483}
  {\bibfield  {journal} {\bibinfo  {journal} {Mon. Not. R. Astron. Soc.}\
  }\textbf {\bibinfo {volume} {91}},\ \bibinfo {pages} {483} (\bibinfo {year}
  {1931})}\BibitemShut {NoStop}%
\bibitem [{\citenamefont {Robertson}(1935)}]{Robertson1935}%
  \BibitemOpen
  \bibfield  {author} {\bibinfo {author} {\bibfnamefont {H.~P.}\ \bibnamefont
  {Robertson}},\ }\bibfield  {title} {\bibinfo {title} {{Kinematics and
  World-Structure}},\ }\href {https://doi.org/10.1086/143681} {\bibfield
  {journal} {\bibinfo  {journal} {Astrophys. J.}\ }\textbf {\bibinfo {volume}
  {82}},\ \bibinfo {pages} {284} (\bibinfo {year} {1935})}\BibitemShut
  {NoStop}%
\bibitem [{\citenamefont {Robertson}(1936{\natexlab{a}})}]{Robertson1936a}%
  \BibitemOpen
  \bibfield  {author} {\bibinfo {author} {\bibfnamefont {H.~P.}\ \bibnamefont
  {Robertson}},\ }\bibfield  {title} {\bibinfo {title} {{Kinematics and
  World-Structure II}},\ }\href {https://doi.org/10.1086/143716} {\bibfield
  {journal} {\bibinfo  {journal} {Astrophys. J.}\ }\textbf {\bibinfo {volume}
  {83}},\ \bibinfo {pages} {187} (\bibinfo {year}
  {1936}{\natexlab{a}})}\BibitemShut {NoStop}%
\bibitem [{\citenamefont {Robertson}(1936{\natexlab{b}})}]{Robertson1936b}%
  \BibitemOpen
  \bibfield  {author} {\bibinfo {author} {\bibfnamefont {H.~P.}\ \bibnamefont
  {Robertson}},\ }\bibfield  {title} {\bibinfo {title} {{Kinematics and
  World-Structure III}},\ }\href {https://doi.org/10.1086/143726} {\bibfield
  {journal} {\bibinfo  {journal} {Astrophys. J.}\ }\textbf {\bibinfo {volume}
  {83}},\ \bibinfo {pages} {257} (\bibinfo {year}
  {1936}{\natexlab{b}})}\BibitemShut {NoStop}%
\bibitem [{\citenamefont {Walker}(1937)}]{Walker1937}%
  \BibitemOpen
  \bibfield  {author} {\bibinfo {author} {\bibfnamefont {A.~G.}\ \bibnamefont
  {Walker}},\ }\bibfield  {title} {\bibinfo {title} {{On Milne's Theory of
  World-Structure*}},\ }\href
  {https://doi.org/https://doi.org/10.1112/plms/s2-42.1.90} {\bibfield
  {journal} {\bibinfo  {journal} {Proc. London Math. Soc.}\ }\textbf {\bibinfo
  {volume} {s2-42}},\ \bibinfo {pages} {90} (\bibinfo {year}
  {1937})}\BibitemShut {NoStop}%
\bibitem [{\citenamefont {Birrell}\ and\ \citenamefont
  {Davies}(1982)}]{Birrell1982}%
  \BibitemOpen
  \bibfield  {author} {\bibinfo {author} {\bibfnamefont {N.~D.}\ \bibnamefont
  {Birrell}}\ and\ \bibinfo {author} {\bibfnamefont {P.~C.~W.}\ \bibnamefont
  {Davies}},\ }\href {https://doi.org/10.1017/CBO9780511622632} {\emph
  {\bibinfo {title} {{Quantum Fields in Curved Space}}}},\ Cambridge Monographs
  on Mathematical Physics\ (\bibinfo  {publisher} {Cambridge University
  Press},\ \bibinfo {address} {Cambridge},\ \bibinfo {year} {1982})\BibitemShut
  {NoStop}%
\bibitem [{\citenamefont {Mukhanov}\ and\ \citenamefont
  {Winitzki}(2007)}]{Mukhanov2007}%
  \BibitemOpen
  \bibfield  {author} {\bibinfo {author} {\bibfnamefont {V.}~\bibnamefont
  {Mukhanov}}\ and\ \bibinfo {author} {\bibfnamefont {S.}~\bibnamefont
  {Winitzki}},\ }\href {https://doi.org/10.1017/CBO9780511809149} {\emph
  {\bibinfo {title} {{Introduction to Quantum Effects in Gravity}}}}\ (\bibinfo
   {publisher} {Cambridge University Press},\ \bibinfo {address} {Cambridge},\
  \bibinfo {year} {2007})\BibitemShut {NoStop}%
\bibitem [{\citenamefont {Kluger}\ \emph {et~al.}(1998)\citenamefont {Kluger},
  \citenamefont {Mottola},\ and\ \citenamefont {Eisenberg}}]{Kluger1998}%
  \BibitemOpen
  \bibfield  {author} {\bibinfo {author} {\bibfnamefont {Y.}~\bibnamefont
  {Kluger}}, \bibinfo {author} {\bibfnamefont {E.}~\bibnamefont {Mottola}},\
  and\ \bibinfo {author} {\bibfnamefont {J.~M.}\ \bibnamefont {Eisenberg}},\
  }\bibfield  {title} {\bibinfo {title} {{Quantum Vlasov equation and its
  Markov limit}},\ }\href {https://doi.org/10.1103/physrevd.58.125015}
  {\bibfield  {journal} {\bibinfo  {journal} {Phys. Rev. D}\ }\textbf {\bibinfo
  {volume} {58}},\ \bibinfo {pages} {125015} (\bibinfo {year}
  {1998})}\BibitemShut {NoStop}%
\bibitem [{\citenamefont {Schmidt}\ \emph {et~al.}(1998)\citenamefont
  {Schmidt}, \citenamefont {Blaschke}, \citenamefont {Ropke}, \citenamefont
  {Smolyansky}, \citenamefont {Prozorkevich},\ and\ \citenamefont
  {Toneev}}]{Schmidt1998}%
  \BibitemOpen
  \bibfield  {author} {\bibinfo {author} {\bibfnamefont {S.~M.}\ \bibnamefont
  {Schmidt}}, \bibinfo {author} {\bibfnamefont {D.}~\bibnamefont {Blaschke}},
  \bibinfo {author} {\bibfnamefont {G.}~\bibnamefont {Ropke}}, \bibinfo
  {author} {\bibfnamefont {S.~A.}\ \bibnamefont {Smolyansky}}, \bibinfo
  {author} {\bibfnamefont {A.~V.}\ \bibnamefont {Prozorkevich}},\ and\ \bibinfo
  {author} {\bibfnamefont {V.~D.}\ \bibnamefont {Toneev}},\ }\bibfield  {title}
  {\bibinfo {title} {{A quantum kinetic equation for particle production in the
  Schwinger mechanism}},\ }\href {https://doi.org/10.1142/S0218301398000403}
  {\bibfield  {journal} {\bibinfo  {journal} {Int. J. Mod. Phys. E}\ }\textbf
  {\bibinfo {volume} {07}},\ \bibinfo {pages} {709} (\bibinfo {year}
  {1998})}\BibitemShut {NoStop}%
\bibitem [{\citenamefont {\'Alvarez-Dom\'inguez}\ \emph
  {et~al.}(2022)\citenamefont {\'Alvarez-Dom\'inguez}, \citenamefont {Garay},\
  and\ \citenamefont {Mart\'in-Benito}}]{Alvarez2022}%
  \BibitemOpen
  \bibfield  {author} {\bibinfo {author} {\bibfnamefont {A.}~\bibnamefont
  {\'Alvarez-Dom\'inguez}}, \bibinfo {author} {\bibfnamefont {L.~J.}\
  \bibnamefont {Garay}},\ and\ \bibinfo {author} {\bibfnamefont
  {M.}~\bibnamefont {Mart\'in-Benito}},\ }\bibfield  {title} {\bibinfo {title}
  {{Generalized quantum Vlasov equation for particle creation and unitary
  dynamics}},\ }\href {https://doi.org/10.1103/PhysRevD.105.125012} {\bibfield
  {journal} {\bibinfo  {journal} {Phys. Rev. D}\ }\textbf {\bibinfo {volume}
  {105}},\ \bibinfo {pages} {125012} (\bibinfo {year} {2022})}\BibitemShut
  {NoStop}%
\bibitem [{\citenamefont {Schmidt}\ \emph {et~al.}(2024)\citenamefont
  {Schmidt}, \citenamefont {Parra-L\'opez}, \citenamefont {Tolosa-Sime\'on},
  \citenamefont {Sparn}, \citenamefont {Kath}, \citenamefont {Liebster},
  \citenamefont {Duchene}, \citenamefont {Strobel}, \citenamefont
  {Oberthaler},\ and\ \citenamefont {Floerchinger}}]{Schmidt2024}%
  \BibitemOpen
  \bibfield  {author} {\bibinfo {author} {\bibfnamefont {C.~F.}\ \bibnamefont
  {Schmidt}}, \bibinfo {author} {\bibfnamefont {A.}~\bibnamefont
  {Parra-L\'opez}}, \bibinfo {author} {\bibfnamefont {M.}~\bibnamefont
  {Tolosa-Sime\'on}}, \bibinfo {author} {\bibfnamefont {M.}~\bibnamefont
  {Sparn}}, \bibinfo {author} {\bibfnamefont {E.}~\bibnamefont {Kath}},
  \bibinfo {author} {\bibfnamefont {N.}~\bibnamefont {Liebster}}, \bibinfo
  {author} {\bibfnamefont {J.}~\bibnamefont {Duchene}}, \bibinfo {author}
  {\bibfnamefont {H.}~\bibnamefont {Strobel}}, \bibinfo {author} {\bibfnamefont
  {M.~K.}\ \bibnamefont {Oberthaler}},\ and\ \bibinfo {author} {\bibfnamefont
  {S.}~\bibnamefont {Floerchinger}},\ }\bibfield  {title} {\bibinfo {title}
  {{Cosmological particle production in a quantum field simulator as a quantum
  mechanical scattering problem}},\ }\href
  {https://doi.org/https://doi.org/10.1103/PhysRevD.110.123523} {\bibfield
  {journal} {\bibinfo  {journal} {Phys. Rev. D}\ }\textbf {\bibinfo {volume}
  {110}},\ \bibinfo {pages} {123523} (\bibinfo {year} {2024})}\BibitemShut
  {NoStop}%
\bibitem [{\citenamefont {Agullo}\ \emph {et~al.}(2024)\citenamefont {Agullo},
  \citenamefont {Delhom},\ and\ \citenamefont
  {Parra-L\'opez}}]{Agullo:2024lry}%
  \BibitemOpen
  \bibfield  {author} {\bibinfo {author} {\bibfnamefont {I.}~\bibnamefont
  {Agullo}}, \bibinfo {author} {\bibfnamefont {A.}~\bibnamefont {Delhom}},\
  and\ \bibinfo {author} {\bibfnamefont {A.}~\bibnamefont {Parra-L\'opez}},\
  }\bibfield  {title} {\bibinfo {title} {{Toward the observation of entangled
  pairs in BEC analog expanding universes}},\ }\href
  {https://doi.org/10.1103/PhysRevD.110.125023} {\bibfield  {journal} {\bibinfo
   {journal} {Phys. Rev. D}\ }\textbf {\bibinfo {volume} {110}},\ \bibinfo
  {pages} {125023} (\bibinfo {year} {2024})}\BibitemShut {NoStop}%
\bibitem [{\citenamefont {Tolosa-Sime\'on}\ \emph {et~al.}(2022)\citenamefont
  {Tolosa-Sime\'on}, \citenamefont {Parra-L\'opez}, \citenamefont
  {S\'anchez-Kuntz}, \citenamefont {Haas}, \citenamefont {Viermann},
  \citenamefont {Sparn}, \citenamefont {Liebster}, \citenamefont {Hans},
  \citenamefont {Kath}, \citenamefont {Strobel}, \citenamefont {Oberthaler},\
  and\ \citenamefont {Floerchinger}}]{Tolosa2022}%
  \BibitemOpen
  \bibfield  {author} {\bibinfo {author} {\bibfnamefont {M.}~\bibnamefont
  {Tolosa-Sime\'on}}, \bibinfo {author} {\bibfnamefont {A.}~\bibnamefont
  {Parra-L\'opez}}, \bibinfo {author} {\bibfnamefont {N.}~\bibnamefont
  {S\'anchez-Kuntz}}, \bibinfo {author} {\bibfnamefont {T.}~\bibnamefont
  {Haas}}, \bibinfo {author} {\bibfnamefont {C.}~\bibnamefont {Viermann}},
  \bibinfo {author} {\bibfnamefont {M.}~\bibnamefont {Sparn}}, \bibinfo
  {author} {\bibfnamefont {N.}~\bibnamefont {Liebster}}, \bibinfo {author}
  {\bibfnamefont {M.}~\bibnamefont {Hans}}, \bibinfo {author} {\bibfnamefont
  {E.}~\bibnamefont {Kath}}, \bibinfo {author} {\bibfnamefont {H.}~\bibnamefont
  {Strobel}}, \bibinfo {author} {\bibfnamefont {M.~K.}\ \bibnamefont
  {Oberthaler}},\ and\ \bibinfo {author} {\bibfnamefont {S.}~\bibnamefont
  {Floerchinger}},\ }\bibfield  {title} {\bibinfo {title} {{Curved and
  expanding spacetime geometries in Bose-Einstein condensates}},\ }\href
  {https://doi.org/10.1103/PhysRevA.106.033313} {\bibfield  {journal} {\bibinfo
   {journal} {Phys. Rev. A}\ }\textbf {\bibinfo {volume} {106}},\ \bibinfo
  {pages} {033313} (\bibinfo {year} {2022})}\BibitemShut {NoStop}%
\bibitem [{\citenamefont {Stwalley}(1976)}]{Stwalley1976}%
  \BibitemOpen
  \bibfield  {author} {\bibinfo {author} {\bibfnamefont {W.~C.}\ \bibnamefont
  {Stwalley}},\ }\bibfield  {title} {\bibinfo {title} {Stability of
  spin-aligned hydrogen at low temperatures and high magnetic fields: New
  field-dependent scattering resonances and predissociations},\ }\href
  {https://doi.org/10.1103/PhysRevLett.37.1628} {\bibfield  {journal} {\bibinfo
   {journal} {Phys. Rev. Lett.}\ }\textbf {\bibinfo {volume} {37}},\ \bibinfo
  {pages} {1628} (\bibinfo {year} {1976})}\BibitemShut {NoStop}%
\bibitem [{\citenamefont {Cornish}\ \emph {et~al.}(2000)\citenamefont
  {Cornish}, \citenamefont {Claussen}, \citenamefont {Roberts}, \citenamefont
  {Cornell},\ and\ \citenamefont {Wieman}}]{Cornish2000}%
  \BibitemOpen
  \bibfield  {author} {\bibinfo {author} {\bibfnamefont {S.~L.}\ \bibnamefont
  {Cornish}}, \bibinfo {author} {\bibfnamefont {N.~R.}\ \bibnamefont
  {Claussen}}, \bibinfo {author} {\bibfnamefont {J.~L.}\ \bibnamefont
  {Roberts}}, \bibinfo {author} {\bibfnamefont {E.~A.}\ \bibnamefont
  {Cornell}},\ and\ \bibinfo {author} {\bibfnamefont {C.~E.}\ \bibnamefont
  {Wieman}},\ }\bibfield  {title} {\bibinfo {title} {{Stable
  ${}^{85}\mathrm{Rb}$ Bose-Einstein Condensates with Widely Tunable
  Interactions}},\ }\href {https://doi.org/10.1103/PhysRevLett.85.1795}
  {\bibfield  {journal} {\bibinfo  {journal} {Phys. Rev. Lett.}\ }\textbf
  {\bibinfo {volume} {85}},\ \bibinfo {pages} {1795} (\bibinfo {year}
  {2000})}\BibitemShut {NoStop}%
\bibitem [{\citenamefont {Chin}\ \emph {et~al.}(2010)\citenamefont {Chin},
  \citenamefont {Grimm}, \citenamefont {Julienne},\ and\ \citenamefont
  {Tiesinga}}]{Chin2010}%
  \BibitemOpen
  \bibfield  {author} {\bibinfo {author} {\bibfnamefont {C.}~\bibnamefont
  {Chin}}, \bibinfo {author} {\bibfnamefont {R.}~\bibnamefont {Grimm}},
  \bibinfo {author} {\bibfnamefont {P.}~\bibnamefont {Julienne}},\ and\
  \bibinfo {author} {\bibfnamefont {E.}~\bibnamefont {Tiesinga}},\ }\bibfield
  {title} {\bibinfo {title} {Feshbach resonances in ultracold gases},\ }\href
  {https://doi.org/10.1103/RevModPhys.82.1225} {\bibfield  {journal} {\bibinfo
  {journal} {Rev. Mod. Phys.}\ }\textbf {\bibinfo {volume} {82}},\ \bibinfo
  {pages} {1225} (\bibinfo {year} {2010})}\BibitemShut {NoStop}%
\bibitem [{\citenamefont {Adorno}\ \emph {et~al.}(2018)\citenamefont {Adorno},
  \citenamefont {Ferreira}, \citenamefont {Gavrilov},\ and\ \citenamefont
  {Gitman}}]{Adorno2018}%
  \BibitemOpen
  \bibfield  {author} {\bibinfo {author} {\bibfnamefont {T.~C.}\ \bibnamefont
  {Adorno}}, \bibinfo {author} {\bibfnamefont {R.}~\bibnamefont {Ferreira}},
  \bibinfo {author} {\bibfnamefont {S.~P.}\ \bibnamefont {Gavrilov}},\ and\
  \bibinfo {author} {\bibfnamefont {D.~M.}\ \bibnamefont {Gitman}},\ }\bibfield
   {title} {\bibinfo {title} {Role of switching-on and -off effects in the
  vacuum instability},\ }\href {https://doi.org/10.1142/S0217751X18500604}
  {\bibfield  {journal} {\bibinfo  {journal} {Int. J. Mod. Phys. A}\ }\textbf
  {\bibinfo {volume} {33}},\ \bibinfo {pages} {1850060} (\bibinfo {year}
  {2018})}\BibitemShut {NoStop}%
\bibitem [{\citenamefont {Hung}\ \emph {et~al.}(2013)\citenamefont {Hung},
  \citenamefont {Gurarie},\ and\ \citenamefont {Chin}}]{Hung2013}%
  \BibitemOpen
  \bibfield  {author} {\bibinfo {author} {\bibfnamefont {C.-L.}\ \bibnamefont
  {Hung}}, \bibinfo {author} {\bibfnamefont {V.}~\bibnamefont {Gurarie}},\ and\
  \bibinfo {author} {\bibfnamefont {C.}~\bibnamefont {Chin}},\ }\bibfield
  {title} {\bibinfo {title} {{From Cosmology to Cold Atoms: Observation of
  Sakharov Oscillations in a Quenched Atomic Superfluid}},\ }\href
  {https://doi.org/10.1126/science.1237557} {\bibfield  {journal} {\bibinfo
  {journal} {Science}\ }\textbf {\bibinfo {volume} {341}},\ \bibinfo {pages}
  {1213–1215} (\bibinfo {year} {2013})}\BibitemShut {NoStop}%
\bibitem [{\citenamefont {Ema}\ \emph {et~al.}(2016)\citenamefont {Ema},
  \citenamefont {Jinno}, \citenamefont {Mukaida},\ and\ \citenamefont
  {Nakayama}}]{Ema2016}%
  \BibitemOpen
  \bibfield  {author} {\bibinfo {author} {\bibfnamefont {Y.}~\bibnamefont
  {Ema}}, \bibinfo {author} {\bibfnamefont {R.}~\bibnamefont {Jinno}}, \bibinfo
  {author} {\bibfnamefont {K.}~\bibnamefont {Mukaida}},\ and\ \bibinfo {author}
  {\bibfnamefont {K.}~\bibnamefont {Nakayama}},\ }\bibfield  {title} {\bibinfo
  {title} {Gravitational particle production in oscillating backgrounds and its
  cosmological implications},\ }\href
  {https://doi.org/10.1103/PhysRevD.94.063517} {\bibfield  {journal} {\bibinfo
  {journal} {Phys. Rev. D}\ }\textbf {\bibinfo {volume} {94}},\ \bibinfo
  {pages} {063517} (\bibinfo {year} {2016})}\BibitemShut {NoStop}%
\bibitem [{\citenamefont {Markkanen}\ and\ \citenamefont
  {Nurmi}(2017)}]{Markkanen2017b}%
  \BibitemOpen
  \bibfield  {author} {\bibinfo {author} {\bibfnamefont {T.}~\bibnamefont
  {Markkanen}}\ and\ \bibinfo {author} {\bibfnamefont {S.}~\bibnamefont
  {Nurmi}},\ }\bibfield  {title} {\bibinfo {title} {Dark matter from
  gravitational particle production at reheating},\ }\href
  {https://doi.org/10.1088/1475-7516/2017/02/008} {\bibfield  {journal}
  {\bibinfo  {journal} {J. Cosmol. Astropart. Phys.}\ }\textbf {\bibinfo
  {volume} {2017}}\bibinfo  {number} { (02)},\ \bibinfo {pages}
  {008}}\BibitemShut {NoStop}%
\bibitem [{\citenamefont {Ema}\ \emph {et~al.}(2018)\citenamefont {Ema},
  \citenamefont {Nakayama},\ and\ \citenamefont {Tang}}]{Ema2018}%
  \BibitemOpen
\bibfield  {number} {  }\bibfield  {author} {\bibinfo {author} {\bibfnamefont
  {Y.}~\bibnamefont {Ema}}, \bibinfo {author} {\bibfnamefont {K.}~\bibnamefont
  {Nakayama}},\ and\ \bibinfo {author} {\bibfnamefont {Y.}~\bibnamefont
  {Tang}},\ }\bibfield  {title} {\bibinfo {title} {Production of purely
  gravitational dark matter},\ }\href
  {https://doi.org/10.1007%2Fjhep09%282018%29135} {\bibfield  {journal}
  {\bibinfo  {journal} {J. High Energ. Phys.}\ }\textbf {\bibinfo {volume}
  {2018}}\bibinfo  {number} { (9)},\ \bibinfo {pages} {135}}\BibitemShut
  {NoStop}%
\bibitem [{\citenamefont {Chung}\ \emph {et~al.}(2019)\citenamefont {Chung},
  \citenamefont {Kolb},\ and\ \citenamefont {Long}}]{Chung2019}%
  \BibitemOpen
\bibfield  {number} {  }\bibfield  {author} {\bibinfo {author} {\bibfnamefont
  {D.~J.~H.}\ \bibnamefont {Chung}}, \bibinfo {author} {\bibfnamefont {E.~W.}\
  \bibnamefont {Kolb}},\ and\ \bibinfo {author} {\bibfnamefont {A.~J.}\
  \bibnamefont {Long}},\ }\bibfield  {title} {\bibinfo {title} {Gravitational
  production of super-hubble-mass particles: an analytic approach},\ }\href
  {https://doi.org/10.1007%2Fjhep01%282019%29189} {\bibfield  {journal}
  {\bibinfo  {journal} {J. High Energ. Phys.}\ }\textbf {\bibinfo {volume}
  {2019}}\bibinfo  {number} { (01)},\ \bibinfo {pages} {189}}\BibitemShut
  {NoStop}%
\bibitem [{\citenamefont {Bastero-Gil}\ \emph {et~al.}(2019)\citenamefont
  {Bastero-Gil}, \citenamefont {Santiago}, \citenamefont {Ubaldi},\ and\
  \citenamefont {Vega-Morales}}]{Bastero2019}%
  \BibitemOpen
\bibfield  {number} {  }\bibfield  {author} {\bibinfo {author} {\bibfnamefont
  {M.}~\bibnamefont {Bastero-Gil}}, \bibinfo {author} {\bibfnamefont
  {J.}~\bibnamefont {Santiago}}, \bibinfo {author} {\bibfnamefont
  {L.}~\bibnamefont {Ubaldi}},\ and\ \bibinfo {author} {\bibfnamefont
  {R.}~\bibnamefont {Vega-Morales}},\ }\bibfield  {title} {\bibinfo {title}
  {Vector dark matter production at the end of inflation},\ }\href
  {https://doi.org/10.1088/1475-7516/2019/04/015} {\bibfield  {journal}
  {\bibinfo  {journal} {J. Cosmol. Astropart. Phys.}\ }\textbf {\bibinfo
  {volume} {2019}}\bibinfo  {number} { (04)},\ \bibinfo {pages}
  {015}}\BibitemShut {NoStop}%
\bibitem [{\citenamefont {Yu}\ \emph {et~al.}(2023)\citenamefont {Yu},
  \citenamefont {Fu},\ and\ \citenamefont {Guo}}]{Yu2023}%
  \BibitemOpen
\bibfield  {number} {  }\bibfield  {author} {\bibinfo {author} {\bibfnamefont
  {Z.}~\bibnamefont {Yu}}, \bibinfo {author} {\bibfnamefont {C.}~\bibnamefont
  {Fu}},\ and\ \bibinfo {author} {\bibfnamefont {Z.-K.}\ \bibnamefont {Guo}},\
  }\bibfield  {title} {\bibinfo {title} {Particle production during inflation
  with a nonminimally coupled spectator scalar field},\ }\href
  {http://dx.doi.org/10.1103/PhysRevD.108.123509} {\bibfield  {journal}
  {\bibinfo  {journal} {Phys. Rev. D}\ }\textbf {\bibinfo {volume} {108}}
  (\bibinfo {year} {2023})}\BibitemShut {NoStop}%
\bibitem [{\citenamefont {Cembranos}\ \emph {et~al.}(2023)\citenamefont
  {Cembranos}, \citenamefont {Garay}, \citenamefont {Parra-L\'opez},\ and\
  \citenamefont {S\'anchez~Vel\'azquez}}]{Cembranos2023}%
  \BibitemOpen
  \bibfield  {author} {\bibinfo {author} {\bibfnamefont {J.~A.}\ \bibnamefont
  {Cembranos}}, \bibinfo {author} {\bibfnamefont {L.~J.}\ \bibnamefont
  {Garay}}, \bibinfo {author} {\bibfnamefont {A.}~\bibnamefont
  {Parra-L\'opez}},\ and\ \bibinfo {author} {\bibfnamefont {J.~M.}\
  \bibnamefont {S\'anchez~Vel\'azquez}},\ }\bibfield  {title} {\bibinfo {title}
  {Late vacuum choice and slow roll approximation in gravitational particle
  production during reheating},\ }\href
  {https://doi.org/10.1088/1475-7516/2023/08/060} {\bibfield  {journal}
  {\bibinfo  {journal} {J. Cosmol. Astropart. Phys.}\ }\textbf {\bibinfo
  {volume} {2023}}\bibinfo  {number} { (08)},\ \bibinfo {pages}
  {060}}\BibitemShut {NoStop}%
\end{thebibliography}%

\end{document}